\documentclass[letterpaper,11pt]{elsarticle}
\pdfoutput=1

\usepackage{graphicx, caption, subcaption}  
\usepackage{dcolumn}   
\usepackage{amsmath, amssymb}        
\usepackage{color,latexsym, array,multirow,  verbatim, enumerate}
\usepackage{url,hyperref}
\usepackage{lineno}

\newcommand{\tev}{~\text{TeV}}
\newcommand{\gev}{~\text{GeV}}

\def\rgj{r^{\gamma/j}}
\def\rjt{r^{j/\tau}}
\def\rgt{r^{\gamma/\tau}}
\def\rab{r^{\alpha/\beta}}

\def\Bgj{\mathcal{B}^{\gamma/j}}
\def\Bjt{\mathcal{B}^{j/\tau}}
\def\Bgt{\mathcal{B}^{\gamma/\tau}}
\def\Bab{\mathcal{B}^{\alpha/\beta}}

\usepackage{geometry}
\hypersetup{
   colorlinks=true,       
   linkcolor=red,        
   citecolor=red,         
   filecolor=magenta,     
  }

\biboptions{sort&compress}
\begin{document}

\title{A Framework for Finding Anomalous Objects at the LHC}
\author[a]{Amit Chakraborty}
\ead{amit@post.kek.jp}
\author[b,c]{Abhishek M.~Iyer}
\ead{iyera@na.infn.it}
\author[c,d]{Tuhin S.~Roy}
\ead{tuhin@theory.tifr.res.in}

\address[a]{Theory Center, Institute of Particle and Nuclear Studies, \\
KEK, 1-1 Oho, Tsukuba, Ibaraki 305-0801, Japan}
\address[b]{INFN-Sezione di Napoli, Via Cintia, 80126 Napoli, Italia}
\address[c]{Department of Theoretical Physics, Tata Institute of Fundamental Research, Homi Bhabha Road, Colaba, Mumbai 400 005, India}
\address[d]{Theory Division T-2, Los Alamos National laboratory, Los Alamos, NM 87545, USA}

\begin{abstract}
Search for new physics events at the LHC mostly rely on the assumption that
the events are characterized in terms of standard-reconstructed objects such as
isolated photons, leptons, and jets initiated by QCD-partons. While such
strategy works for a vast majority of physics beyond the standard model
scenarios, there are examples aplenty where new physics give rise to anomalous
objects (such as collimated and equally energetic particles, decays due to long
lived particles etc.) in the detectors, which can not be classified as any of
the standard-objects. Varied methods and search strategies have been proposed,
each of which is trained and optimized for specific models, topologies, and
model parameters. Further, as LHC keeps excluding all expected candidates for
new physics, the need for a generic method/tool that is capable of finding the
unexpected can not be understated. In this paper, we propose one such method
that relies on the philosophy that all anomalous objects are \emph{not} 
standard-objects. The anomaly finder, we suggest, simply is a collection of
vetoes that eliminate all standard-objects up to a pre-determined acceptance
rate. Any event containing at least one anomalous object (that passes all these
vetoes), can be identified as a candidate for new physics. Subsequent offline
analyses can determine the nature of the anomalous object as well as of the
event, paving a robust way to search for these new physics scenarios in a
model-independent fashion. Further, since the method relies on learning only
the standard-objects, for which control samples are readily available from
data, one can build the analysis in an entirely data-driven way.

\end{abstract}





\maketitle
\flushbottom


\newpage
\section{\label{sec:1}Introduction}
The discovery of the Higgs boson of the Standard Model (SM) of particle physics  in the Large Hadron Collider (LHC) 
\cite{Aad:2012tfa,Chatrchyan:2012xdj}, was believed to be a precursor towards the realization of non-standard physics at around the TeV scale. However, the analysis of all data from Run-I and  Run-II so far have failed to yield any statistically significant excess over the SM expectations in any of the channels being looked at \cite{Sirunyan:2017fsj,CMS:2017qjo,CMS:2017sqn,ATLAS:2017tmd,Aaboud:2017dmy,ATLAS:2017uun}. While it is highly likely that new physics (NP) is just around the corner and is going to show up as LHC keeps accumulating data, it is worthwhile to think through whether there remains gaps in aspects of our search strategies where events due to NP might show up and yet elude our grasp. 

However, before proceeding further, let us deconstruct the general search strategy being employed at the LHC. Broadly speaking, at the detector level events due to collisions  are recorded  in terms of the charged tracks observed at the trackers and the muon spectrometers, energies deposited at different cells of the electromagnetic calorimeters (namely, ECAL), and the hadronic calorimeters (namely, HCAL). The CMS collaboration of the LHC employs a sophisticated particle-flow algorithm~\cite{Beaudette:2014cea} which combines all this information and generates outputs as a set of $4$-vectors, which are then classified into objects such as electrons, muons, photons, charged and neutral hadrons. Note that, these particle-flow objects, even though carry names of the particles, should still be treated as detector objects since further processing is required before one can start the process of identifying the physics of short-distance that might have given rise to the event. 

The \textit{detector-objects} (either tracks and calorimeter cells or even  particle-flow objects) are the inputs to a series of algorithms and techniques that are used to obtain the \textit{reconstructed objects} such as isolated photons, electrons, muons, taus and jets\footnote{These jets, often understood to be initiated by hard partons from short distance physics, which undergoes further showering and hadronization  are usually thought to be synonymous with `QCD-jets'.}. An event is now described in terms of these `standard' reconstructed objects along with some variables that carry the global detector information such as, missing energy, $H_T$ etc. Standard phenomenological studies to search for NP as well as SM physics at the LHC use this information. 

The above-mentioned strategy works fairly well for the SM and a large fraction of NP physics processes. However, the fundamental assumption that \emph{all} NP events can be described in terms of these reconstructed objects is not true. Take, for example, reconstructed photons -- these are outputs of an algorithm which identifies a cluster of ECAL energy depositions to be a photon if the pattern of energy deposits is consistent with the shower of a photon in the calorimeter \cite{MITREVSKI20162539,Khachatryan:2015iwa}. However, it is not implausible to imagine a NP scenario which gives rise to \emph{only} collimated photons (known as photon-jets~\cite{Dobrescu:2000jt,Chang:2006bw,Toro:2012sv,Draper:2012xt,Ellis:2012sd,Ellis:2012zp,
Agrawal:2015dbf,Fukuda:2016qah,Knapen:2015dap,Chang:2015sdy,Curtin:2013fra,Dasgupta:2016wxw,Allanach:2017qbs}) instead of single photons, where the degree of collimation is less than the size of a reconstructed photon. In this case, the photon-finder algorithm, trained on the samples of showers from single photons may not find any photon in the event. As a result, either we completely miss the event or, at best, the event gets classified as an event consisting of QCD-jets. Photon jets are not the only example -- one can again find such examples where NP gives rise to events consisting of `anomalous' or `non-standard' objects, such as collimated electrons (or, electron-jets~\cite{Aad:2012qua,Aad:2014yea}, or, say lepton jets in general \cite{Falkowski:2010gv,Falkowski:2010cm,Ruderman:2009tj,Iyer:2016qzd,Barello:2016zlb,Dube:2017jgo,Cheung:2009su,
Chang:2016lfq,Buschmann:2015awa}), collimated taus (or tau-jets~\cite{Katz:2010iq,Englert:2011iz}), particle with large life-times (e.g., long lived 
particles~\cite{Evans:2016zau,Allanach:2016pam,Aad:2015rba,ATLAS:2016jza,CMS:2014bra,Aad:2015asa,Aaij:2014nma,CMS:2014wda,Dev:2016vle,Ito:2017dpm,Banerjee:2017hmw}), etc. to name a few.

Several methods and search strategies have been proposed, trained, and optimized to find many of these scenarios by identifying these anomalous objects. An essential problem remaining is that these strategies are powerful when it comes to finding specific NP models and topologies for which the searches have been optimized, but lose sensitivity fast, when models/topologies/parameters are varied.  In other words, no general framework exists to probe and trigger these events with anomalous objects at the LHC. In this paper we attempt to provide one such general framework that can be used to select (and store) these events containing  anomalous objects (equivalently signatures of NP) for further physics analysis.    

The framework proposed here relies on the broad definition of anomalous objects, as objects that are \emph{not} standard such as photons, electrons, taus, or QCD-jets.  The philosophy is, therefore, straightforward -- understand the standard-objects enough to be able to veto these at a desired level of efficiency. The objects that pass through these series of vetoes are, therefore, anomalous.  The working principle can be briefly summarized as follows: 
\begin{enumerate}
\item First, we find reconstructed-objects by clustering all the calorimeter information, using a \emph{single} algorithm and a \emph{single} set of clustering parameters (this conforms with the philosophy first proposed in Ref.~\cite{Ellis:2012zp,Ellis:2012sd}). The output then becomes the superset of all standard as well as anomalous objects. Additionally, we demand that these outputs satisfy certain hardness criteria, which ensures that these objects can not be resultants from noise only.
    
\item Using a set of judiciously chosen variables, we find representations of these reconstructed-objects in a multi-dimensional space. By training MultiVariateAnalyses (MVAs) we identify patches in this multi-dimensional space occupied by the standard objects (namely, single photons, single electrons, single tau (hadronic), and QCD-jets). 
\item Finally, we construct vetoes that simply block these patches rich in standard objects. In quantitative terms, these vetoes require `target-rates', defined as the rates at which standard-objects will be acceptable. For example, if one sets the target-rate for QCD-jets to be $1\%$, this in turn determines the veto-boundary such that only $1\%$ of QCD-jets can pass it. 
\item Objects that pass through these vetoes are then identified as anomalous objects. Events containing at least one anomalous object become candidates for events due to NP and need to be recorded. 
One can look at the multidimensional representation of an anomalous object (offline) to learn about the object itself (such as whether it contains collimated photons, or it corresponds to long-lived objects, etc.). Coupled with the event information (such as the number, the nature, and the kinematic features of the accompanying objects in the event), one can then identify whether the event arises from NP or from SM. 
\end{enumerate}
The crucial feature of this strategy is that the whole exercise relies on knowing standard objects, such as single photons, single electrons, single taus, QCD-parton initiated jets etc., for which we have ample data that can work as control samples. Therefore, the entire formalism can be easily turned into a data-driven exercise, even though, in this paper we rely on Monte Carlo in order to demonstrate its working principle. Furthermore, this framework has plenty of 
rooms to improve, since it offers flexibility in terms of easily including new variables.  We also emphasize that, even though, standard objects such as isolated photons, electrons, etc. can be subsets of outputs in the first step,  we are not proposing any new method/changes in the way these standard objects are identified currently.  Rather, we propose that this procedure be implemented in parallel to current strategies, and be used only to identify the presence of anomalous objects in the event.

The paper is organized as follows: in Sec.\ref{sec:2} we outline the working principle and the philosophy of the proposed framework; in Sec.\ref{sec:3} we discuss an ensemble of jet-variables that we employ in order to construct the veto; in Sec.\ref{sec:4} we demonstrate the construction of vetoes,  using responses of carefully constructed MVAs; in Sec.\ref{sec:5} we give examples, where anomalous objects manage to pass these vetoes at acceptable rates (in particular, we give examples of collimated photons, electrons, and taus) even though vetoes did not use any information pertaining to these anomalous objects; and finally in Sec.\ref{sec:6} we conclude.

\section{\label{sec:2}The philosophy and the Framework}

As mentioned in the introduction, the aim of this paper is to construct a \emph{tool} or a  \emph{methodology}  that can identify an ``anomalous" or ``non-standard"  object, where the adjective anomalous or non-standard refers to the fact that the chance for the chosen object to be a standard object (such as $e,\gamma,\tau$ or QCD-jet) is  highly unlikely (statistically speaking). The fundamental feature of the tool that we attempt to build  is that it can be designed/optimized in an entirely data-driven procedure, even though in this work we use Monte Carlo in order to construct a complete proposal as well as to demonstrate its efficacy. This constraint is non-trivial, since we can not expect to have controlled samples of anomalous objects available at the LHC. 

The aim of this section is to discuss the philosophy of this paper along with its blueprint. This lays the groundwork before we move on and describe the procedures in detail in the following sections. 

\subsection{\label{sec:2.1}A universal framework for analyzing all objects} 

A difficulty arises while implementing such an analysis  is the fact that the ``standard-objects" are reconstructed objects. Even though the experimental analysis reconstructs these using the same detector elements such as calorimeter cells and tracks, or more refine objects such as particle flow elements, however varied reconstruction algorithms and/or parameters are used to find different objects. This makes a direct comparison among different reconstructed objects somewhat ambiguous. A robust analysis needs a universal construct for all objects ``standard" or ``non-standard", built from calorimeters and trackers. In this work we implement a formalism as proposed in Ref.~\cite{Ellis:2012sd,Ellis:2012zp}. The key ingredient is that one uses `jets',  defined as the output of a standard Infrared (IR) safe jet algorithm, to be the common construct for all physics objects that deposit energy in the calorimeters. 

Note that the formalism adopted here maintains a clear distinction between the terminology of `jets' and `QCD-jets'. We define `jets' as the output of IR safe jet algorithms such as anti-$k_T$ \cite{Cacciari:2008gp}, $k_T$ \cite{Ellis:1993tq,Catani:1993hr}, or C/A \cite{Dokshitzer:1997in}, which, in some instances, may have nothing to do with partons in QCD. A jet, therefore, becomes a generic concept that is defined in terms of the energy deposits in calorimeter cells and is identified by a jet algorithm. With this definition a QCD-jet is simply a special kind of jet (or rather, a standard-jet). The set of jets, therefore, also includes clustered energetic cells due to a single photon, or an electron, or a tau. 

Our next strategy would be to devise a set of chosen variables in order to identify/classify the jets into categories. The working principle behind this is simple: the variables pave a way to map a jet to a point in a multi-dimensional space; a potent set of variables can ensure that jets of different kinds cluster in different corners in the space; as a consequence, by identifying these corners one can tag photons/electrons/taus/QCD-jets at the same time while minimizing the mistag rates due to jets of other kinds.  It turns out that  jet substructure techniques~\cite{Thaler:2010tr,Thaler:2011gf,Larkoski:2013eya},  developed to distinguish QCD-jets from jets containing boosted heavy particle decays by probing in detail the energy distribution within the jet, are ideal for this job. In fact, this treatment has been demonstrated to yield higher tagging efficiency for photons for the same mistag rate due to QCD-jets. Additionally, this method imparts the advantage of using grooming techniques~\cite{Butterworth:2008iy,Krohn:2009th,Ellis:2009me} in photon tagging, making the tagging performance to be more pile-up robust. 

Refs.\cite{Ellis:2012sd,Ellis:2012zp} also show that the same treatment can be used to find jets consisting of energetic and collimated photons (also known as photon-jets).  Since kinematic features of the underlying physics (\textit{e.g.}, the masses and spins of intermediary particles, whose decay give rise to these objects) are responsible for these distributions, the existence of structures within photon-jets is guaranteed.  Substructure variables, therefore, should be efficient at finding and discriminating photon-jets from QCD-jets and even from single photons. 

In this paper we use a slightly altered philosophy. In Refs.\cite{Ellis:2012sd,Ellis:2012zp}, the authors rely on understanding photon-jets in order to separate these from single photons and QCD-jets. The analysis was more focused to obtain the best signal acceptance rate through performing a signal-background optimization 
procedure using several jet observables in a MVA framework. The analysis in Ref \cite{Ellis:2012zp}, like any 
other supervised learning, is extremely powerful in discriminating the photon-jets 
from QCD-jets. However, this technique quickly looses its discrimination power if, for example, 
photon-jets are replaced by ditau jets, or even use collimated photons but produced with 
different kinematics. Thus, the analysis of \cite{Ellis:2012sd,Ellis:2012zp}, though extremely useful, uses knowledge on the type of NP and thereby limited to the specific new physics scenario 
under consideration. In this paper, however, we use a slightly altered philosophy; we follow the 
`unsupervised learning' technique. We start with various standard objects (electron, photon, tau and 
QCD samples), while being completely agnostic of the type of NP, and go on understanding various properties of 
each of these standard objects. We then systematically construct vetoes to identify regions of phase space where the standard jets have small acceptance rate. As a result, jets that escape these vetos, will, to a high probability, be considered as `non-standard' objects, and corresponding events will be triggered as potential candidates for new physics events. 

It is to be stressed that while constructing the vetoes only the known 
properties of the standard jets are used, no new physics input has been considered here. The proposed 
framework is thus less powerful compared to the one obtained after supervised 
learning that discriminates a specific kind of non-standard object from the standard objects, for example \cite{Ellis:2012sd,Ellis:2012zp}, however, is more powerful in terms of its applicability in finding wide varieties of non-standard objects, and, therefore, can be used as a universal trigger for probing new physics signatures at the LHC.

\subsection{\label{sec:2.2} Standard-jets}
The second step towards constructing vetoes is to learn about the standard-jets. In this work we focus on  four kinds of standard-jets, namely photons, electrons, taus (hadronic), and QCD-jets. The purpose of  this subsection is to outline the operational definitions of these objects. The details of event generation, object reconstruction and the involved pile-up analysis are discussed in \ref{sec:appA}.
\begin{itemize}
\item \textbf{Photons:}  We cluster the calorimeter responses for the events $pp \rightarrow h \rightarrow  \gamma \gamma$ using anti-$k_T$ jet algorithm for $R=0.4$ and $p_T > 50\gev$. From each event only the hardest jet, obtained after performing a pile-up subtraction, is selected. In order to create a pure sample of jets initiated by photons, we impose an additional consistency criterion using Monte Carlo (MC) truth. We check that the selected jet indeed contains at least one energetic photon inside. To be specific, there should be at least one MC photon within $\Delta R < 0.4$ from the jet axis, where the angular separation between two four-vectors is defined via $\left(\Delta R\right)^2 \equiv \left( \Delta\eta \right)^2 + \left( \Delta\phi \right)^2$. The quantities  $\Delta\eta$ and $\Delta\phi$, refer to 
the differences in pseudo-rapidity and azimuthal angle of the two four-vectors respectively. 
From now on each jet of this sample will be known as a jet of type photon, or a jet initiated by a photon  (or, often simply a photon or $\gamma$).
 
\item \textbf{Electrons:} The simplest and the most practical choice  is to use jets initiated by electrons from the decay $Z \rightarrow e e$.  In this work, however, we use electrons from Monte Carlo sample where Higgs is being used as the intermediate particle in order to generate samples. We have explicitly checked that the distributions of the substructure variables we employ here remain identical irrespective of whether we use $Z$ or $h$ as the intermediate particle. To be specific, we cluster the calorimetric responses for the events  $pp \rightarrow h \rightarrow  e e$  using the anti-$k_T$ jet algorithm with $R=0.4$ and $p_T > 50\gev$. We then select the leading jet, obtained after performing a pile-up subtraction, from each event as long as it also contains at least one MC electron within $\Delta R < 0.4$. We call jets from this  sample to be a jet of type electron or a jet initiated by an electron (or, often simply an electron or $e$). 
  
\item \textbf{Taus:} Similar to the case of electrons, the most practical choice is to have jets initiated by taus from decays $Z \rightarrow \tau \tau$. However, we simulate the events $pp \rightarrow h \rightarrow  \tau^{+} \tau^{-}$ with the $\tau$ decaying hadronically. The jets are then constructed using anti-$k_T$ jet algorithm with jet radius $R = 0.4$ and $p_T > 50\gev$. Similar to the earlier cases, the leading jet from each event, obtained after performing a pile-up subtraction, is selected as long as there is at least one MC tau within $\Delta R < 0.4$.  We denote each jet from this sample as a jet of type tau or a jet initiated by tau (or, often simply as a tau or $\tau$).

\item \textbf{QCD-jets:} Hard QCD processes are simulated with a minimum $p_T$ threshold of $50\gev$. Jets are then constructed from the calorimetric four-vectors using anti-$k_T$ jet algorithm with $R=0.4$ and $p_T > 50\gev$. For each event, the leading ($p_T$ ordered) jet obtained after performing a pile-up subtraction, is selected for further analysis. We require no further purity criteria for these jets. We denote the jets from this sample as jets of type QCD-jets or  jets initiated by QCD-partons or simply QCD-jets or simply as $j$.
\end{itemize} 


Before concluding this subsection, let us discuss two important issues: first, the choice of jet radius $R$ = 0.4
and second, the use of Higgs boson as the intermediate particle. In a typical search for boosted massive resonances, the jet radius $R$ is
chosen such that the resultant jet contains (almost) all the decay products of the resonance. The search
strategy then needs to customize $R$ by optimizing the discovery potential of the target resonance. The problem we are solving here is
unconventional; we do not have any particular target resonance mass in mind. By aiming at those
cases where the angular separation among the decay products is such that the standard techniques fail,
we get a target $R$ -- namely $R$ needs to be smaller than (or, at most equal to) the
size of the standard reconstructed objects ($\sim$ 0.4). In fact, given the choice of the new
physics model under consideration (see \ref{sec:appA}), we find the choice of $R$ = 0.4 includes all the
decay products of the collimated objects, and therefore it's already a very robust choice. Therefore,
increasing $R$ will not improve signal acceptance, however, it will necessarily increase hadronic contaminations of 
the underlying events and pile-ups. In such a case, QCD-jets need to be controlled 
separately to improve the sensitivity of the new physics. We use the Higgs scalar as the intermediate
particle to generate standard-jets only for convenience. During implementation, we rather recommend
the use of $Z$ for generating electrons and taus. For example, leading jets in events with
di-boson (namely, $ ZZ \rightarrow 4e$) can be used to populate the electron sample.

\subsection{\label{sec:2.3}From substructure variables to a veto}

The next agenda on the list is to construct a veto for all standard objects by looking at these objects only. We do this in multiple stages: 
\begin{enumerate}
\item  Using a carefully chosen set of variables, a jet is mapped to a point in a multidimensional space. To elaborate, one can translate the statement such as ``mass of a jet (say, $J$) is $m_J$" to the statement that the variable mass maps $J \mapsto m_J$. Following the same logic, we use a set of variables $\{ V_1, V_2, \dots , V_D \} $, to map each jet $J$ to a set of numbers $\{ v_1, v_2, \dots , v_D \} $.  Assigning the jet $J$ a vector of numbers  $\vec{v}  \equiv \{ v_1, v_2, \dots , v_D \} $, one finds a representation of the jet $J$ in the $D$-dimensional space. 

\item We use Greek indices to denote the type of jets. In particular, if a standard-jet is designated as $J_\alpha$, then $\alpha$ represents one of $\gamma, e, \tau$ or $j$.  A set of variables, therefore, maps the $i$-th jet of kind $\alpha$ (namely, $J_{\alpha, i}$) to a representation $\vec{v}_{\alpha, i}$.
  
\item As we noted before, the variables are chosen in such a manner that one can simply find corners (or close regions) in the $D$-dimensional space where the standard-jets occupy and use $D$-dimensional boxes to isolate these samples. However, as $D$ increases the analysis simply becomes tedious and less and less manageable. 

\item In order to overcome the difficulty mentioned above, we incur a 
mechanism that maps the $D$-dimensional vector of numbers $\vec{v}$ to a vector of fewer numbers, while still keeping jets of different types separated from each other.  To be specific we use MultiVariateAnalyses, in particular, Boosted Decision Tree or BDT in the ROOT framework~\cite{Speckmayer:2010zz} (see \ref{sec:appA} for BDT specific parameter details.). The process can be described as follows: 
\begin{enumerate}[i.]

\item The input to a BDT is jets of two kinds with a set of variables that are, ideally, efficient in discriminating  these two jets.  As explained before, this set of variables give jets their representations. The job for  the BDT is, therefore, to separate a list of jets of type $\alpha$ (or, the set of vectors $\{ \vec{v}_{\alpha} \}$) from another list of   jets of type $\beta$ (or, the set of vectors $\{ \vec{v}_{\beta} \}$). 

\item Broadly speaking, the BDT optimizes the separation of jets, by dividing the multidimensional space in many hyper-boxes, which are dominantly populated by jets of one kind in an algorithmic way. Now, given any point in this multi-dimensional space, a BDT can associate with it a response that is calculated based on the hyper-boxes that the point belongs to, as well as the purity contents of each box. Once a BDT is successfully trained to separate signals from backgrounds, it assigns large responses for signal-like jets whereas small responses to background-like jets. We denote a BDT treating  jets of type $\alpha $ to be signal like, and jets of type $\beta$ to be background like, by $\Bab$ and its responses by $\rab$.  We rescale the responses such that, the distribution for responses for jets of type-$\alpha$ (namely, $\rab_\alpha$) peaks at large values (close to $1$), whereas the same for jets of type-$\beta$ (namely, $\rab_\beta$) peaks at smaller values (close to $0$).

\item Summarizing, a BDT optimized to separate jets of type $\alpha$ from type $\beta$ (represented by $\Bab$),  maps any jet  $J$ (represented by a vector $\vec{v}_J$),  to a response (a number) $r_J^{\alpha/\beta}$. 
\begin{equation} 
J \ \xrightarrow{\ \{ V_1, V_2, \dots , V_D \} \ }  \ \vec{v}_J    \ \xrightarrow{\ \Bab \ } 
\  r_J^{\alpha/\beta} \; . 
\end{equation} 
As explained before, we expect $\rab_\alpha$ close to $1$,  whereas $\rab_\beta$ close to $0$. There is no definite prediction for any other kind of jets (except that we expect it to be somewhere between $0$ and $1$).  
\end{enumerate}
The advantage of the above procedure is straightforward. Even if more and more variables are added to the existing set $ \{ V_1, V_2, \dots , V_D \} $, the jet still gets mapped to a single number for a  BDT. 

\item In this work, we end up using three BDTs ($ \Bjt$, $ \Bgt$, and $\Bgj$), and therefore map all jets to a point in the $\left( \rjt, \rgt, \rgj  \right)$ space.  The entire procedure reduces a $D$-dimensional  representation to a $3$-dimensional representation without sacrificing information pertaining to pair-wise differences between the standard-jets. 

\item As we show later, by construction, standard-jets occupy rather small corners in this space. Finally, after identifying bins in these three dimensions rich in  standard-jets, we can veto most of standard-jets. 

\end{enumerate}

\subsection{Summary} 
\begin{itemize}
\item We attempt to devise a tool which identifies anomalous objects, defined as the objects that are not the standard-objects such as electrons, photons, taus, and QCD-jets. The procedure therefore is synonymous to the construction of vetoes that block these objects.  

\item The fundamental problem in comparing all of these standard or anomalous objects is that we need a universal construct. For this purpose, we employ IR safe jet algorithms whose output (namely, jets) become the common construct.  Electrons, photons, taus, and QCD-jets are therefore jets of specific types, so as all anomalous objects. 
  
\item We represent jets by points in a $D$-dimensional space spanned by outputs of $D$-number of jet variables. A judicious choice of variables is needed that emphasizes the differences among the jets of different types. 

\item For the vetoes to be effective, we need $D$ to be large which makes the construction of vetoes hard. Increasing $D$, even by $1$, only increases the difficulty associated with the procedure exponentially. We use MVAs (in particular, BDTs) that collapses $D$-dimensional representations to $3$-dimensional representations of the responses. By construction, this reduction of dimensionality  preserves  information pertaining to pair-wise differences between the standard-jets. 

\item As a result, standard-jets get maximally separated from each other in this space. We block these corners rich in standard-jets to construct vetoes.  

\end{itemize}

\section{\label{sec:3}The Variables}

In this section we describe the list of variables which can be useful in characterizing a jet of a given type. The variables 
are based on the tracker and calorimeter information, and also take into account the information associated to the constituents 
of the jets.

\begin{figure}[!htb]
	\begin{center}
		\begin{tabular}{cc}
			\includegraphics[width=7.2cm]{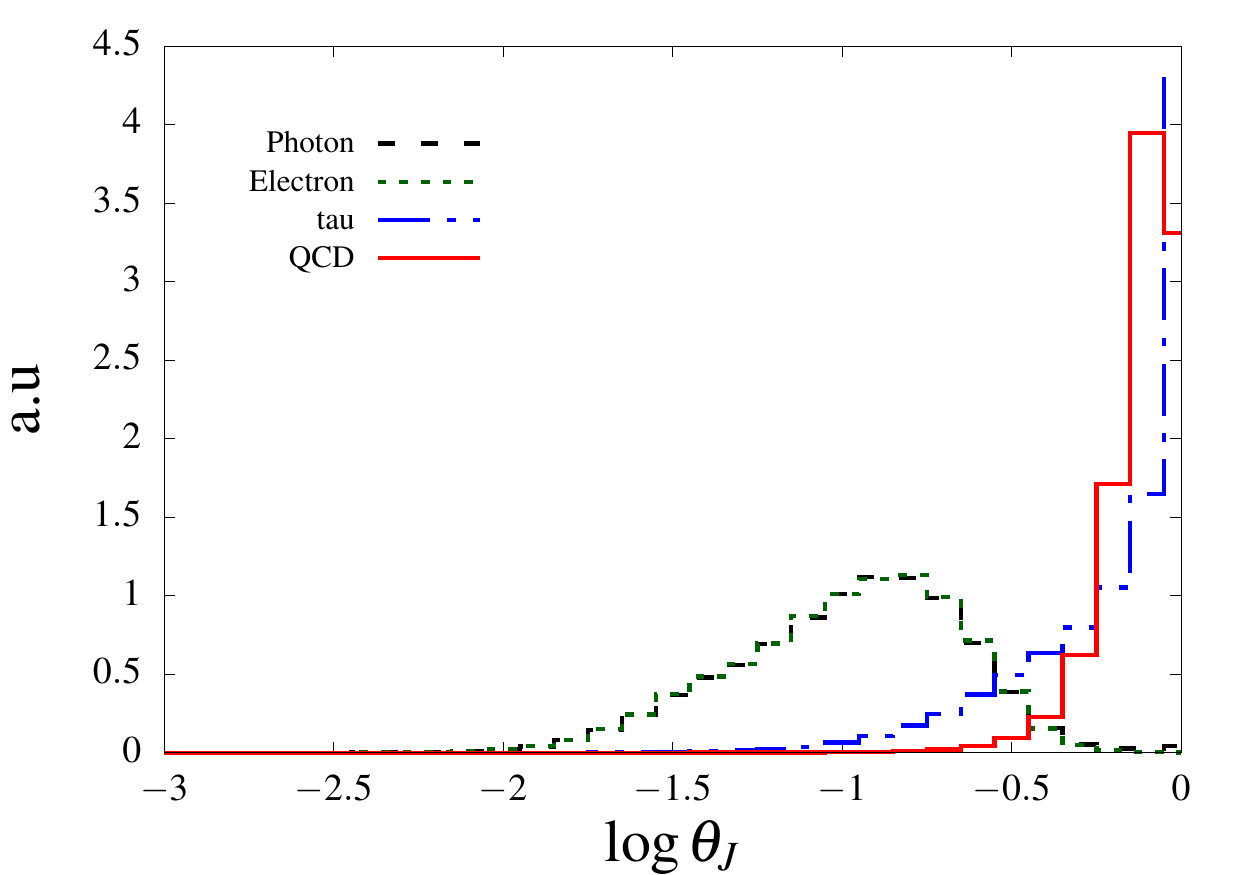}&
			\includegraphics[width=7.2cm]{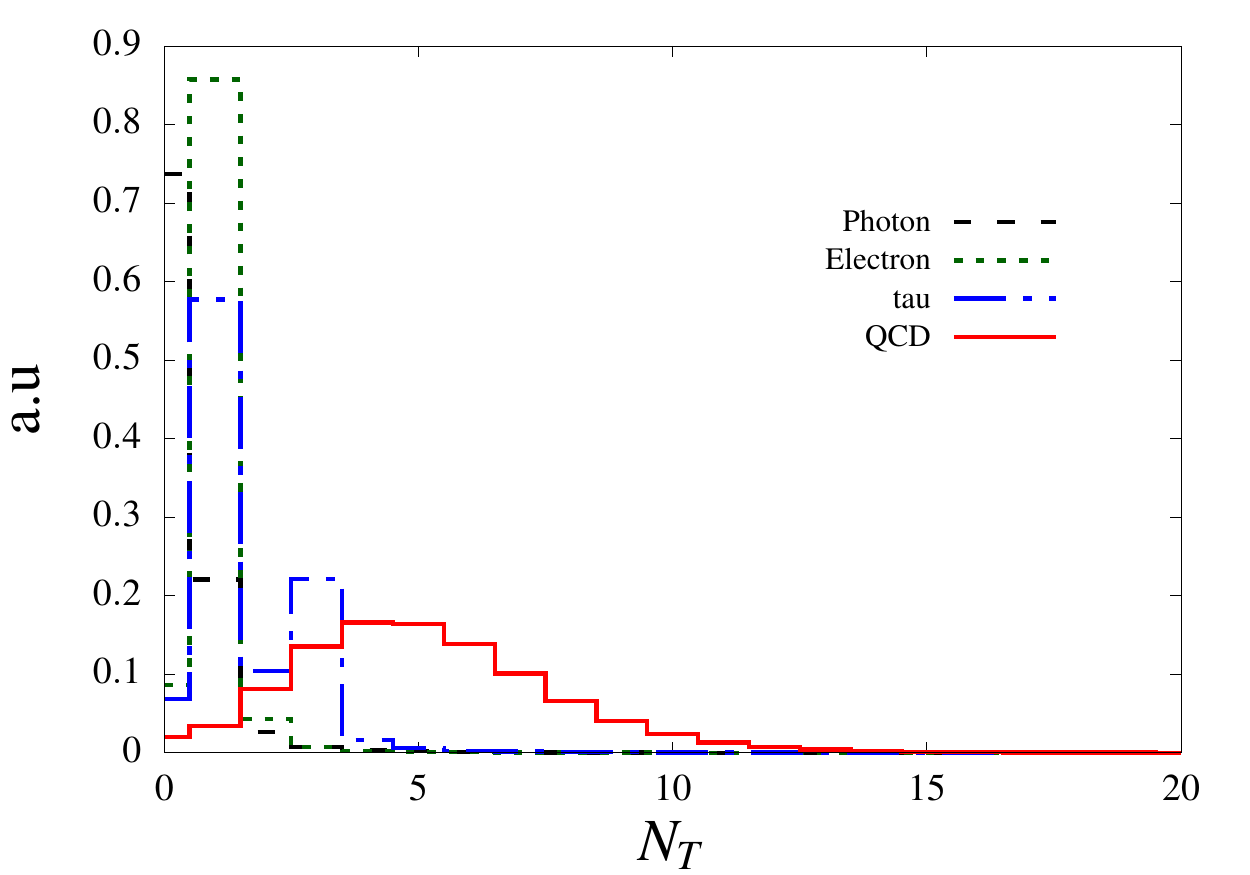}
		\end{tabular}
	\end{center}
	\caption{ \it{The distributions of Hadronic energy fraction (left) and the 
number of tracks (right) in the leading jet for the standard objects.}}
	\protect\label{thetaJ&NT}
\end{figure} 

\subsection{\label{sec:3.2.1} Hadronic energy fraction (namely, $\theta_J$)} 
Since we construct jets from the calorimeter towers, calculating the hadronic energy fraction is particularly easy. Given a jet, we define its hadronic energy function from its constituents, which are calorimeter cells by definition. 
\begin{equation}
\theta_J \ \equiv \ \frac{1}{E_J}  \sum_{i \in \{\text{J, HCAL} \} } E_i \quad \text{ where } \quad  
E_J \ \equiv \ \sum_{i \in J} E_i \, .
\end{equation}
In the above definitions, the sum runs over all constituents of the jet. The total energy of the jet is therefore given by $E_J$.

The $\log \theta_J$ distributions for various kinds of background jets (or say `standard objects') are shown in the left panel of Fig.\ref{thetaJ&NT}. As expected, $\theta_J$ peaks at $1$ for $\tau$-jets, since it dominantly decay to charged pions which deposit almost entire energy in the hadronic calorimeter. On the other hand, QCD-jets contain a significant number of neutral pions ($1/3$ on average because of isospin symmetry) which decay to pair of photons, and thus $\theta_J$ peaks at a smaller value. However, the electron and photon initiated jets deposit almost all their energy in the electromagnetic calorimeter leading to much smaller values of $ \log \theta_J$.  Not surprisingly,  $\theta_J$ is widely used for providing pure samples of electrons and photons. Precise prediction of these distributions for standard objects helps us to understand and probe the presence, if any, of non-standard objects in an event. We are thus going to use this variable extensively in our analysis.   

\subsection{\label{sec:3.2.1}Tracks (namely, $N_T$)} 
The number of tracks  associated to a jet is a measure of charged particle multiplicity inside a jet. Since the multiplicity of particles (charged or not) inside a jet is IR-unsafe, we set a lower $p_T$ threshold and accept tracks which satisfy $p_T > 2\gev$. The number of tracks in the leading jet is counted by calculating $\Delta R$ between the leading jet and each pile-up subtracted track. We then accept those tracks which satisfy $\Delta R <0.4$, where $\left(\Delta R\right)^2 \equiv \left( \Delta\eta \right)^2 + \left( \Delta\phi \right)^2$ with $\Delta\eta$ and $\Delta\phi$ being the differences in pseudo-rapidity and azimuthal angle of the jet and the given track respectively.

The $N_T$ distributions for each kind of background jets are shown in the right panel of Fig.\ref{thetaJ&NT}. A QCD jet or  a jet initiated by colored partons (quarks or gluons) is mostly characterized by a large number of charged particles (\textit{i.e.,} a large $N_T$). These charge particles are mostly hadrons, generated in the hadronization of partons after the initiating parton showers and split into multiple partons. In Fig.\ref{thetaJ&NT} the $N_T$ distribution is peaked around $5$. Note that this value of peak is a function of the size of the jet (\textit{i.e.}, the $R$ parameter in jet clustering), and the minimum value of $p_T$ of the tracks. The distribution moves to the right if $R$ is increased or if the cut on track $p_T$ is lowered. Also note that the $N_T$ distribution depends on the flavor of the parton initiating jets, and often are used for discriminating quark/gluon initiating jets~\cite{Brodsky:1976mg, Gallicchio:2010sw, Gallicchio:2011xc, Gallicchio:2011xq,Gallicchio:2012ez,Bhattacherjee:2015psa,Bhattacherjee:2016bpy}. 

Among the rest of the background jets, photons peaks at zero, while electrons and $\tau$-jets dominantly peak around unity. The $\tau$-jet samples also have a fair amount of three track events due to three charged pions. Because of conversion of photons into charged particles inside the tracker, some of the photons appear in $N_T=1$ bin. We outline the details of photon conversions as implemented in our simulation in \ref{sec:appA}.


\subsection{\label{sec:3.2.1}Energy-momentum distribution in subjets}
In order to quantify the energy-momentum distributions among the subjets of a given jet, we recluster its constituents using $k_T$ algorithm 
\cite{Ellis:1993tq,Catani:1993hr} such that all constituent $4$-vectors are combined and reproduces the original jet $4$-vector. Even though, the final jet $4$-vector remain the same, this procedure assigns the jet a new clustering history. Using this procedure of reclustering the constituents, one can assign a $k_T$ ordered clustering history to any jet irrespective of the jet-algorithm used to find the jet. After reclustering, we obtain exclusive $k_t$-subjets. Of course, the number of exclusive subjets $n_t$ is a free parameter. We then order these subjets according to their transverse momenta such that the subjet momenta follow the relation $p_{T_i} > p_{T_j}$ for $j>i$, with the $0$-th subjet being the hardest.  We primarily concentrate on two variables:  the first one quantifies the fraction of the jet energy (or rather the $p_T$) carried by the leading subjet (namely, $\lambda_J$), while the second variable contains additional information of the next-to-leading as well as next-to-next-to-leading jets (namely, energy-energy correlation or $\epsilon_J$). 
\begin{equation}
\begin{split}
\lambda_J \ \equiv \ \log\left(1-\frac{p_{T_0}}{p_{T_J}}\right) \\
\epsilon_J \ \equiv \ \frac{1}{E_J^2} \sum_{n_f>i>j} E_i E_j
\end{split}
\label{eq:lambda}
\end{equation}
where, as explained before, $p_{T_i}, E_{i}$ is the transverse momentum and energy of the $i$-th subjet (ordered in $p_T$, such the  $0$-th subjet is the hardest); $p_{T_J}, E_{J}$ is transverse momentum and energy of the given jet; and $n_f$ is less or equal to the total number of exclusive subjets ($n_t$) of the given jet. In this work, following Ref.~\cite{Ellis:2012sd,Ellis:2012zp}, we ask for $n_t =5$ and $n_f=3$.  

\begin{figure}[htb!]
	\begin{center}
		\begin{tabular}{cc}
			\includegraphics[width=7.2cm]{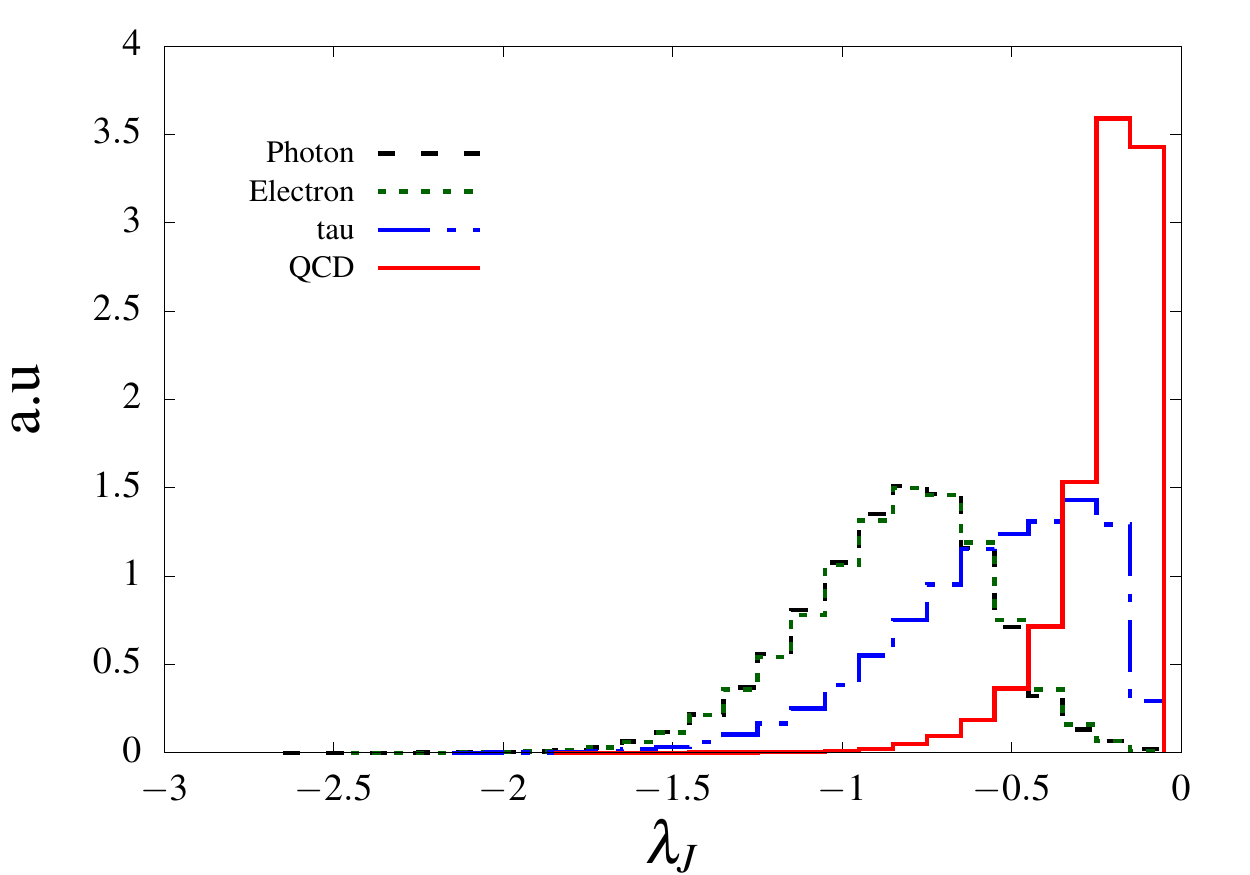}&\includegraphics[width=7.2cm]{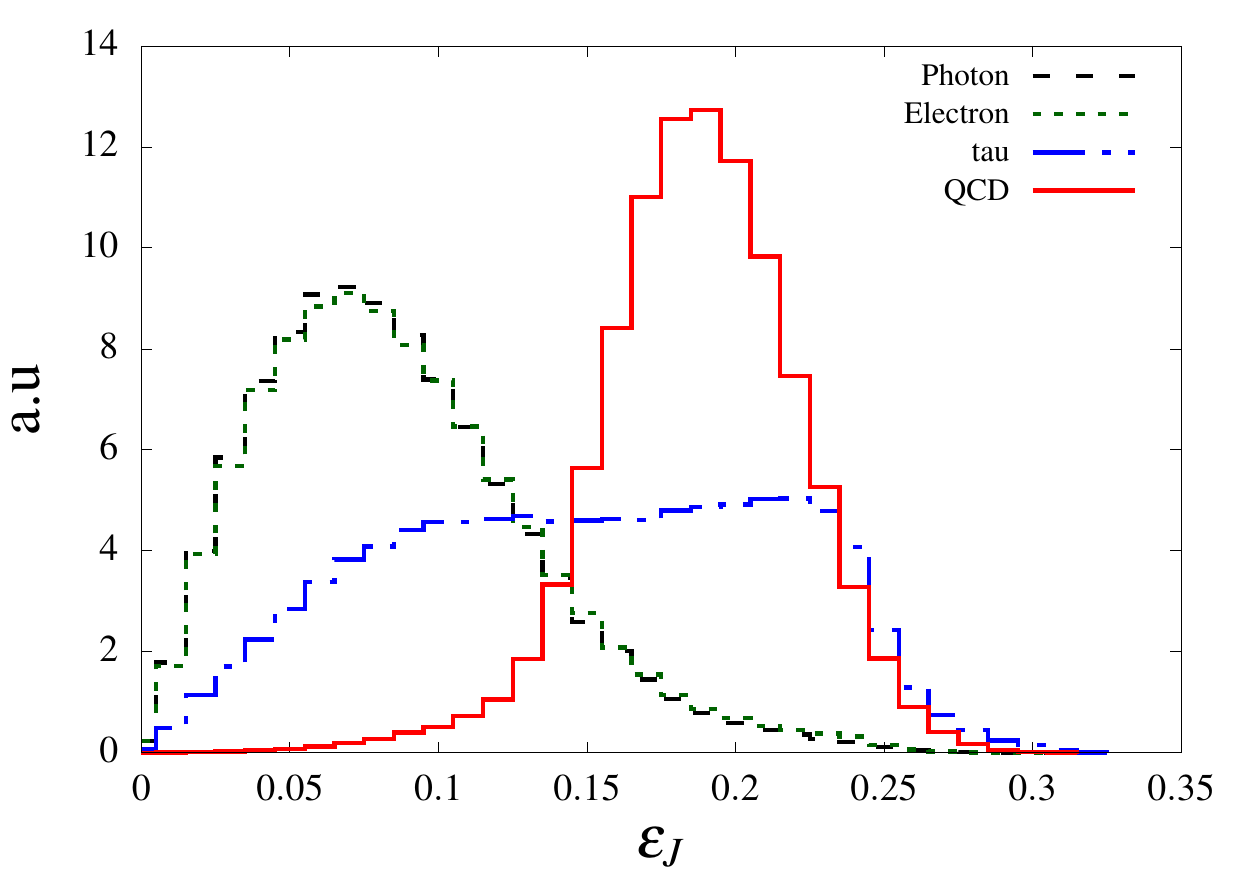}
		\end{tabular}
	\end{center}
	\caption{\it{The distributions of $\lambda_J$ (left) and $\epsilon_J$ (right) for the standard objects.}} 
	\protect\label{LambdaJ}
\end{figure}

For a narrow pencil like (i.e., single prong) jet, the leading subjet carries most of the energy. For these jets one typically gets $p_{T_L} \lesssim p_{T_J}$, and consequently small $\lambda_J$ and $\epsilon_J$.  To be specific,  consider a jet consisting of $n$-number of energetic subjets, then by definition we have the following inequalities: 
\begin{equation}
p_{T_0} \geq p_{T_J}/n \qquad \implies \qquad \lambda \leq \log \left( 1- \frac{1}{n} \right)  
\quad
\implies \quad  \lambda \rightarrow  \left\{ 
\begin{aligned}
-\infty  & \quad \text{as }  n \rightarrow 1 \\
0  & \quad \text{as }  n \rightarrow \infty
\end{aligned} 
\right. \; ,
\label{eq:lambda-n}
\end{equation} 
where we have used the notations as used previously in Eq.~\eqref{eq:lambda}.   Note that  for $n=2, 3, 4, \dots$, one obtains  $\lambda = -0.30, -0.18, -0.12, \dots $ respectively.  As a result, a cut on $\lambda$ is straightforward to understand and interpret. For example, for a jet with $\lambda > -0.30$, the leading subjet contains less than $50\%$ of the total $p_T$. One can intuit from the above fact that the jet most likely contains at least two energetic subjets. Similarly, a jet with $\lambda > -0.18$, most likely will be characterized by three energetic subjets. Therefore, a cut $\lambda > -0.18$, for example, typically allows jets with three or more prongs.  

Similar qualitative understanding can be obtained for $\epsilon_J$.  For example, if we assume the leading jet carries 90\% of the jet energy, then the remaining 10\% will be distributed among other subjets. In that case, the $\epsilon_J$ is expected to be around 0.08-0.09. However, if we assume that the energy distribution among the leading and two sub-leading jets are 50\%, 30\% and 20\% of the total jet energy respectively, then we expect $\epsilon_J$ to be around 0.3. As the number of subjets increases with equal share of energies, $\epsilon_J$ increases.  For $e$ or $\gamma$ initiated jets we expect the distributions of $\lambda_J$ to be peaked at lower values than the QCD jets. Such intuitions are validated in Fig.\ref{LambdaJ}, where we plot $\lambda_J$ (left) and $\epsilon_J $ (right) for all the standard objects. From Fig.\ref{LambdaJ} and the discussion above, it is evident that $\lambda_J$ and $\epsilon_J$ are qualitatively similar in describing the substructure of a given jet. A cut on $\lambda_J$ can be mapped to a corresponding cut in $\epsilon_J$, thereby exhibiting a strong correlation between the two.

\subsection{\label{sec:3.2.2}$N$-subjettiness (namely, $\tau_N$) }
$N$-subjettiness~\cite{Thaler:2010tr} is a measure of the number of energetic subjets (or energy lobes) \emph{inside a jet} as opposed to $N$-jettiness~\cite{Stewart:2010tn} which is an example of an event shape. We compute $\tau_N$ of the given jet using the definition in  Ref.~\cite{Thaler:2010tr}. Given a set of $N$-axes, one defines   
\begin{equation}
\tau_N \ \equiv \ \frac{\sum_k p_{T_k}\times \text{min}\left( \Delta R_{1k},\Delta R_{2k}\ldots \Delta R_{Nk}\right)}{\sum p_{T_k} \times R } \; , \qquad \text{ and } \qquad   \tau_{ab} \ \equiv \ \tau_a/\tau_b \; ,
\label{nsubjettiness}
\end{equation}
where $k$ runs over the constituents of the jet,  $p_{T_k}$ is the transverse momentum for the $k$-th constituent, $\Delta R_{ak}$ is the angular distance between the $k$-th constituent and the $a$-th axis. Further, in order to calculate $\tau_N$, one needs $N$-axes. In this work, we use axes collinear to the $N$ exclusive $k_t$-subjets of the jets.  Finally,  Eq.~\eqref{nsubjettiness} also gives the notation for the ratio of two $N$-subjettiness. 
 
In order to understand the physics of $N$-subjettiness, consider for example a jet with $l$ number of distinct lobes of energy. If one calculates $\tau_N$ as a function of $N$ starting with $N=1$, one finds that $\tau_N$ keeps decreasing with increasing $N$, with the rate of decrease maximized around $N=l$. The jet with $l$ prongs, is then characterized by a large drop $\tau_{l-1} \gg \tau_l$.  We can therefore use the ratio variable  $\tau_{N(N-1)}$ to identify the energy distribution inside jet. In an ideal scenario, a jet with $l$ prongs, will be given by a small $\tau_{N(N-1)}$ for $N=l$.  We also find that it is  often useful to consider the product of ratios $\tau_{a(a-1)} \times \tau_{b(b-1)}$, in order to isolate mixed samples containing primarily jets with $a$ or $b$ number of distinct prongs. 

\begin{figure}[!htb]
	\begin{center}
		\begin{tabular}{cc}
			\includegraphics[width=7.2cm]{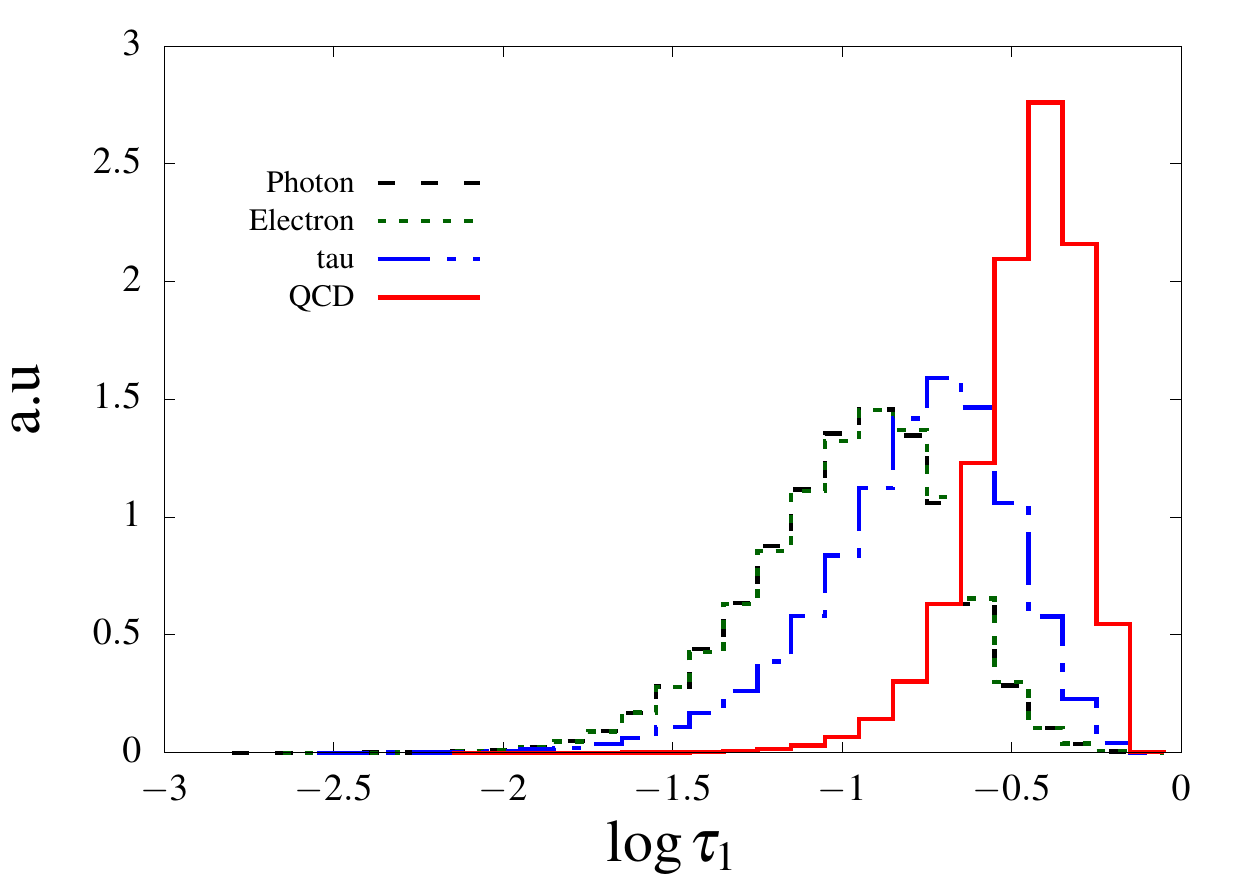}&\includegraphics[width=7.2cm]{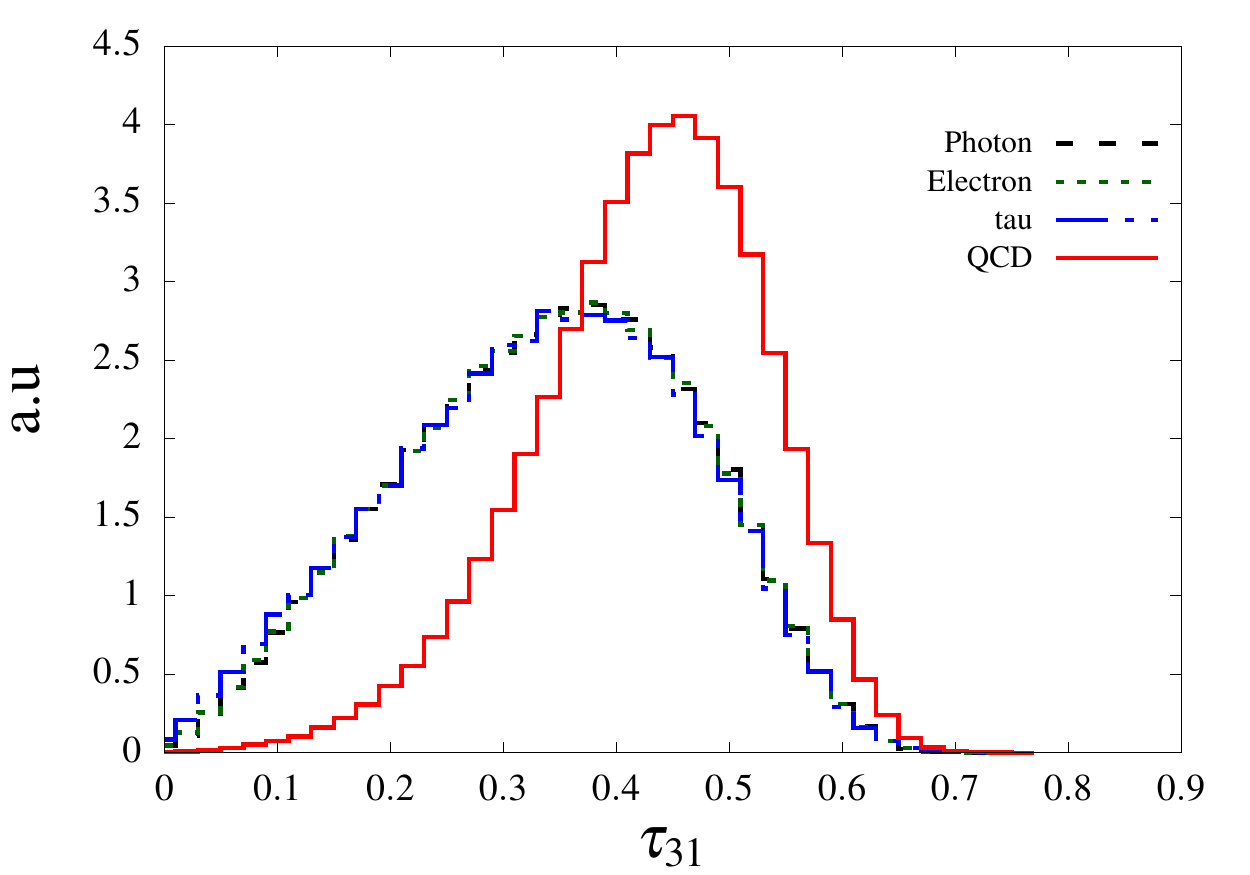}
		\end{tabular}
	\end{center}
	\caption{\it Distribution of $N$-subjettiness variables $\log(\tau_1)$ (left) and $\tau_{31}$ (right) for all 
the standard objects.}
	\protect\label{tau1}
\end{figure}

Out of various possible $\tau_N$ and the ratios $\tau_{ab}$, we  find $\tau_1$ and $\tau_{31}$  particularly to be interesting. In the left panel of Fig.\ref{tau1}, we display the $\log(\tau_1)$ estimated for various background jets.  Jets with energy distributed in a single and narrow prong (such an $e$ or a $\gamma$ initiated jet), is characterized by a small $\tau_1$, whereas jets with broader distributions of energy (such as jets due to QCD) will give rise to sizable $\tau_1$s.  From the left panel one can also, see that $\tau$ initiated jets lie in-between  the parton-initiated jets and the $e/\gamma$-initiated jets, since these are still ``cleaner" than the qcd-jets. In fact some of the $\tau$-initiated jets are characterized by a single pencil like distribution of energy as one sees with  $e/\gamma$-jets. 
The $\tau$- jets lie in between the $e,\gamma$ and QCD jets as they either exhibit a $1$- or $3$- pronged structure. For the latter case, $\tau_1>>0$ and thus has a reasonable overlap with the $QCD$ jets. 
In the right plot of  Fig.\ref{tau1}, we also show the distribution of $\tau_{31}=\frac{\tau_3}{\tau_1}$, which complements the $\log(\tau_1)$ distribution. Since a QCD jet exhibits a broader distribution of energy, it is likely to have multiple prongs inside the jet. As a result, $\tau_3$ may not be significantly smaller to $\tau_1$. For the $e,\gamma$ jets however, $\tau_3$ is much smaller in comparison and is reflected in the plot. The $\tau$- jets are characterized  by $\tau_1(\tau_3)\rightarrow 0$ corresponding to a 1-(3-) pronged structure. Thus the ratio behaves similar to the pencil like jets of $e,\gamma$.

\subsection{\label{sec:3.2.3}Energy Correlation functions and their ratios }
Similar to $N$-subjettiness, energy correlation functions (namely $e_N$) also quantify the distribution of energy inside a jet. The key difference is that the $N$-subjettiness is constructed using the $p_T$ of the constituents weighted by their angular distances from a set of axes, whereas in the definitions of $e_N$,  the weighing parameters are the angles between the constituents themselves.  In particular, we use the following \cite{Larkoski:2013eya,Moult:2016cvt},  
\begin{equation}\label{eq:ecf}
e_N \ = \ \sum_{i_1<i_2<\ldots < i_N\in J}  z_{i_1} z_{i_2} \dots z_{i_N}  \ 
{\left(\prod_{b=1}^{N-1}\prod_{c=b+1}^N \Delta R_{i_bi_c}\right)}^{\beta} \; ,
\quad \text{where} \quad z_i \equiv \frac{p_{T_i}} {p_{T_J}} \; .
\end{equation}
In the equation above the sum runs over all constituents of the jet, and we assume the angular exponent ($\beta$) to be equal to unity. Note that in order to construct  $e_N$ we use dimensionless quantity  $z_i$, which describes the fraction of the jet's transverse momentum carried by its $i$-th constituent.  Consequently $e_N$ is dimensionless. Also note that $e_0$ is taken to be equal to be $1$.  Additionally, we also use correlation ratios and double ratios (ratios of ratios): 
\begin{equation}
\begin{split}
r_N \ &= \ \frac{e_{N+1}} {e_{N}}  \\
C_N \ &= \ \frac{r_{N}} {r_{N-1}} \ = \ \frac{e_{N+1} e_{N-1}}{e_{N}^2}.  
\end{split}
\end{equation}
Understanding the correlation functions and their ratios are straightforward. For a jet with $n$ distinct pencil-like structures, it is clear that there can at maximum be $n$-number of subjets, where all are separated from each other by large angles. Therefore, $e_{n+1}$ is suppressed w.r.t $e_n$. Both ratios and double ratios are sensitive to this fact. The double ratio, in fact, can be employed to measure the higher-order  radiation from leading order substructure. 
\begin{figure}[htb!]
	\begin{center}
		\begin{tabular}{cc}
			\includegraphics[width=7.2cm]{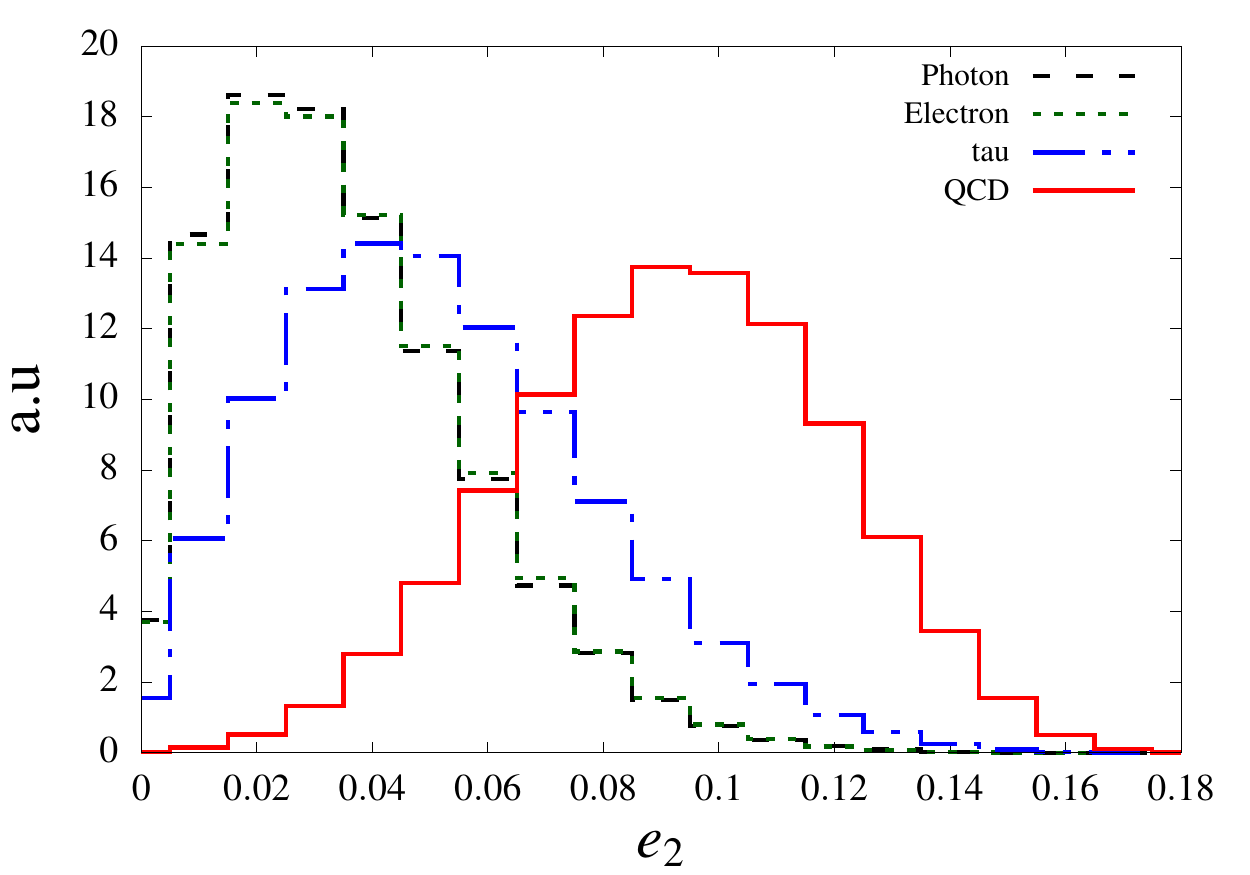}&\includegraphics[width=7.2cm]{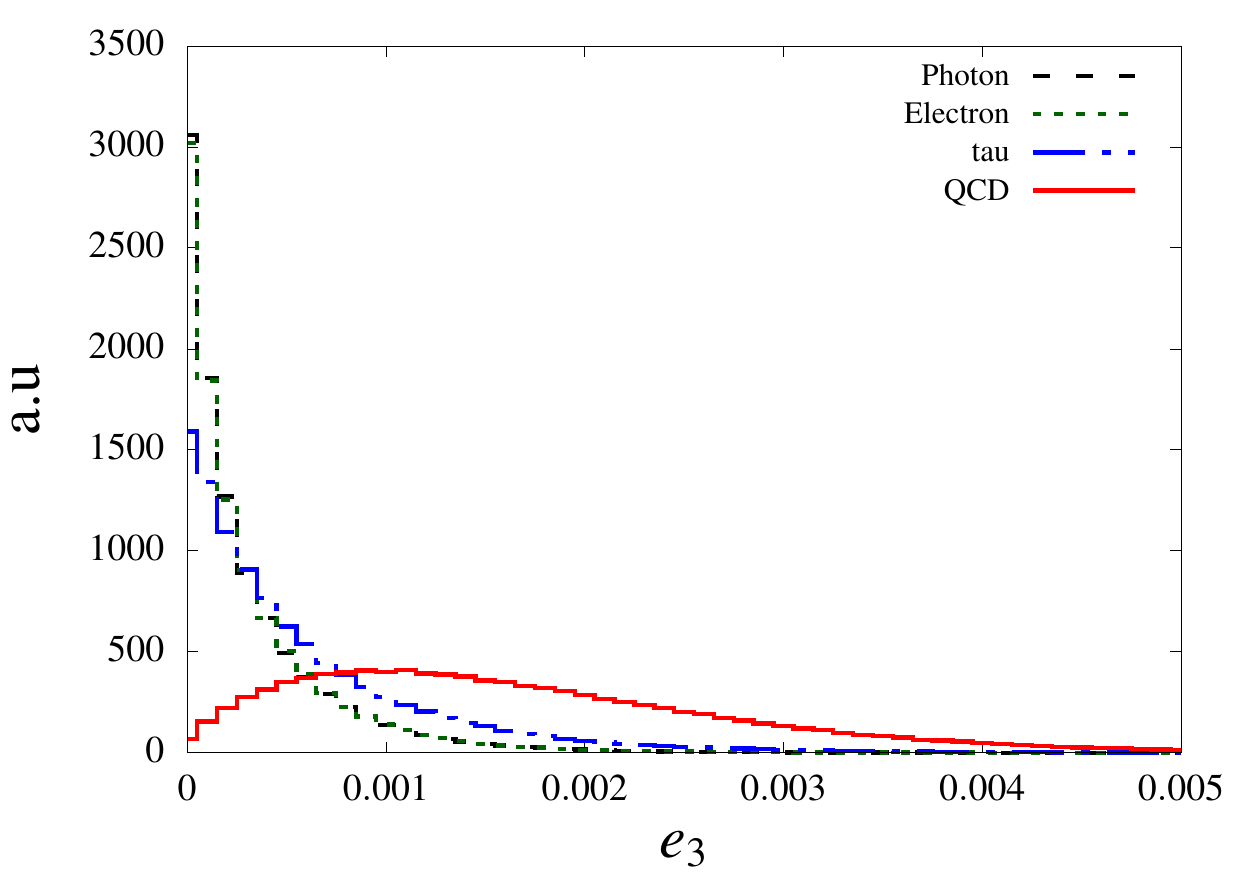}\\
		\end{tabular}
	\end{center}
	\caption{\it Distribution of the Energy Correlation Function (ECF) variables $e_2$ (left) and $e_3$ (right) for all the standard objects. }
	\protect\label{ecf1}
\end{figure}

\begin{figure}[htb!]
	\begin{center}
		\begin{tabular}{cc}
			\includegraphics[width=7.2cm]{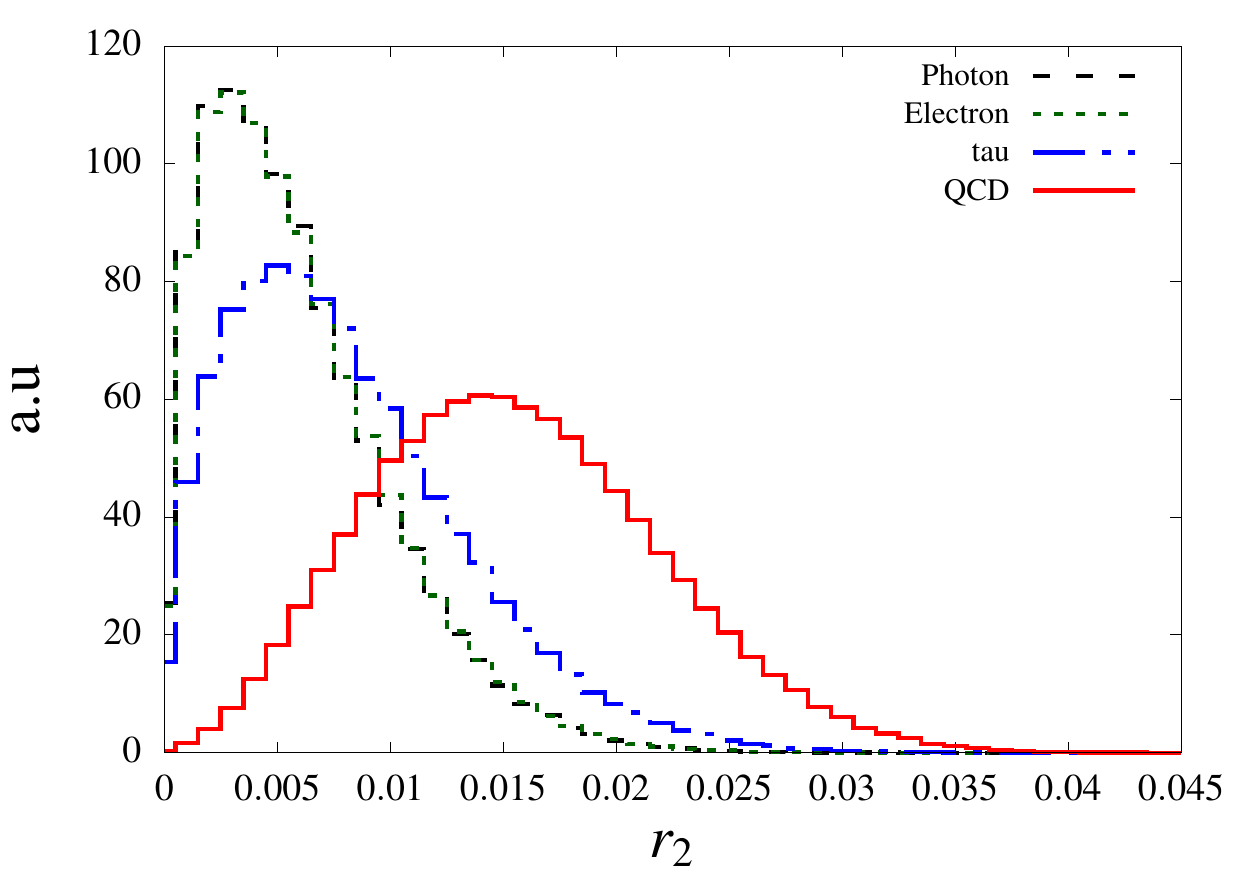}&\includegraphics[width=7.2cm]{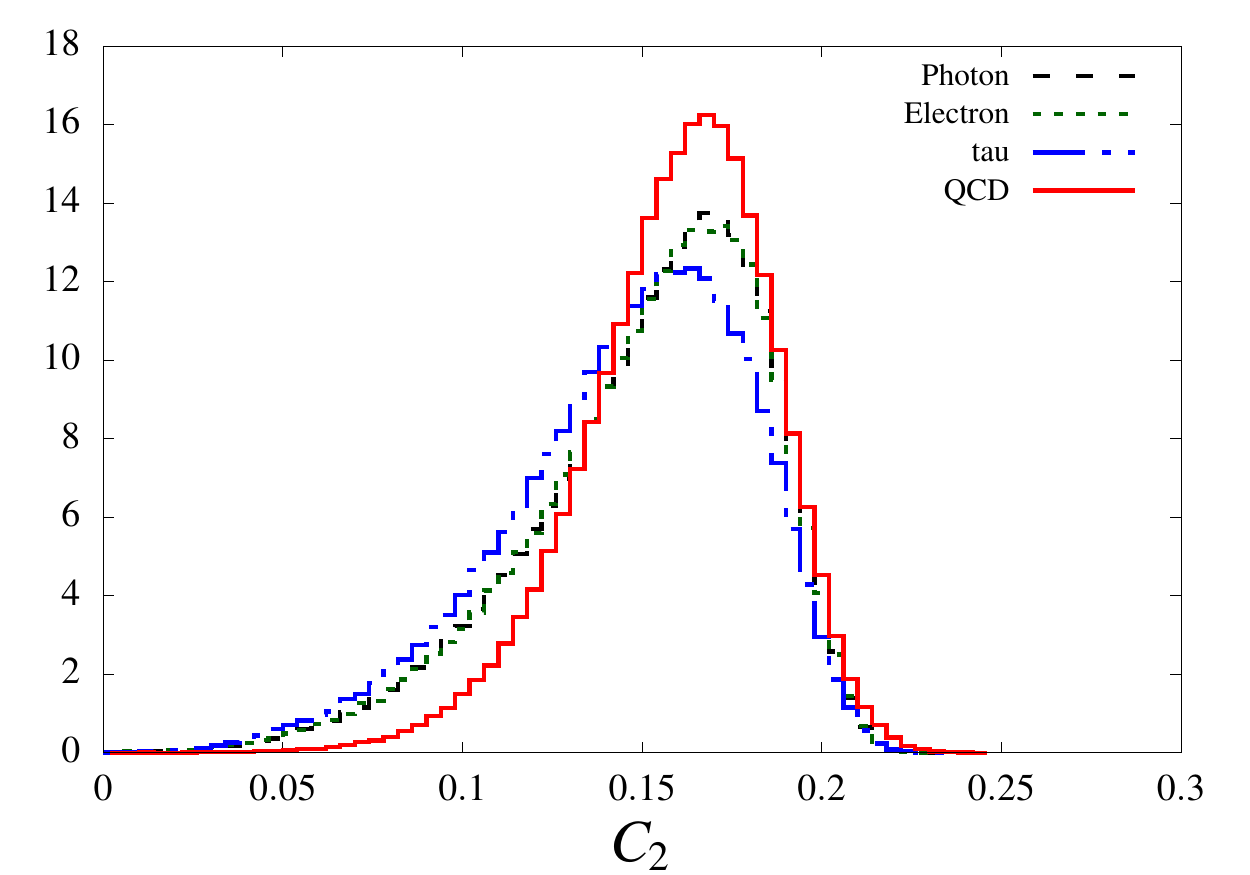}\\
		\end{tabular}
	\end{center}
	\caption{\it Distribution of one of the ratios and double ratios of Energy Correlation Function (ECF) $r_2$ (left) and $C_2$ (right) respectively for all the standard objects. }
	\protect\label{ecf2}
\end{figure}

In Fig.\ref{ecf1}, we show the distribution of the two ECFs, namely $e_2$ and $e_3$, while Fig.\ref{ecf2} displays the 
distribution of the variables involving the ratios of the ECFs. As we have already discussed, for single prong objects like single 
electron, single photon and signal tau we expect both $e_2$ and $e_3$ to be sufficiently small. However, for QCD-jets, being multi-prong 
structure, both $e_2$ and $e_3$ can be large enough. Thus, we expect the distributions of $r_2 \equiv \frac{e_3}{e_2}$ will be shifted towards 
left for the $e, \gamma, \tau$ and right shifted for the QCD-jets. Similar behavior can be seen in $C_2$, however the separation is not 
so significant as it involves a ratio $\frac{e_1}{e_2}$ which is comparable for all of these standard objects.


\section{\label{sec:4}From substructure variables to a veto: a demonstration }

The purpose of this paper is to provide a simple example where we design a relatively simple veto to discard all standard-jets. In the previous section we have summarized a set of variables and for each of these we have examined the distributions of jets of various kinds. As explained before, after examining  the distributions $\{ \vec{v}_\alpha \}$,  we can identify the patches in the multidimensional space which predominantly get occupied by jets of kind $\alpha$. We can simply block these patches in order to veto standard-jets. Even though the procedure seems simple, difficulties arise because of the large number of variables -- one needs to be clever. 

Note that the variables discussed in the last section are all efficient in highlighting differences among jets of different types. However, two among these, namely $\theta_J$ and $N_T$ are special. These are the easiest to comprehend and at the same time, no other variables separate different jets as efficiently as these two. In our analysis, we will first employ these two variables to separate the phase space into many segments (see Subsec.~\ref{sec:4.1}).  In Subsec.~\ref{sec:4.2}, we proceed to analyze those different segments by constructing a realistic veto using multivariate analysis.   
\subsection{\label{sec:4.1} Segmentation of phase space }

Schematically, we segment jets first binning according to their electromagnetic characters and then further binning using the number of  associated tracks. The arguments are simple:  jets with $\theta < \theta_0$ is rich with electromagnetic radiation (mostly neutral pions), and is less likely to be initiated from partons. The count of tracks is also a fairly good indicators of the origin of the jet. Small track multiplicities (small charged hadron multiplicities) indicate small particle multiplicities overall in the jets, which makes them unlikely to be due to  QCD partons. It is then clear that even the use of simple variables such as  $\theta_J$, and $N_T$ can already generate these patches where these are primarily occupied by standard-jets of distinct types. 
\begin{table}[htb!]	
	\begin{center}
		\begin{tabular}{|c||c|c||c|c||c|c||c|c|}
			\hline
			&	
			\multicolumn{2}{|c||}{$\epsilon_\gamma$ (in $\%$)} &
			\multicolumn{2}{|c||}{$\epsilon_e$ (in $\%$)} &
			\multicolumn{2}{|c||}{{$\epsilon_\tau$ (in $\%$)}} &
			\multicolumn{2}{|c|}{{$\epsilon_j$ (in $\%$)}} \\
			\cline{2-9}
			&	
			$\theta < \theta_0$ & $\theta \geq \theta_0$ &  
			$\theta < \theta_0$ & $\theta \geq \theta_0$ &  
			$\theta < \theta_0$ & $\theta \geq \theta_0$ &  
			$\theta < \theta_0$ & $\theta \geq \theta_0$  	\\
			\cline{2-9}
			\multirow{2}{*}{}& 
			\multirow{2}{*}{93.1}&\multirow{2}{*}{6.9}	& 
			\multirow{2}{*}{93.9}&\multirow{2}{*}{6.1}	& 
			\multirow{2}{*}{10.8}&\multirow{2}{*}{89.2}	& 
			\multirow{2}{*}{0.9}&\multirow{2}{*}{99.1}	 	\\ &&&&&&&&\\
			\hline		
			\multirow{2}{*}{$N_T  = 0$ }& 
			\multirow{2}{*}{69.7}&\multirow{2}{*}{3.9}	& 
			\multirow{2}{*}{8.0}&\multirow{2}{*}{0.64}	& 
			\multirow{2}{*}{1.6}&\multirow{2}{*}{5.3}	& 
			\multirow{2}{*}{0.15}&\multirow{2}{*}{1.9}	 	\\ &&&&&&&&\\
			
			\multirow{2}{*}{$N_T  = 1$ }& 
			\multirow{2}{*}{20.6}&\multirow{2}{*}{1.4}	& 
			\multirow{2}{*}{81.3}&\multirow{2}{*}{4.5}	& 
			\multirow{2}{*}{8.5}&\multirow{2}{*}{49.2}	& 
			\multirow{2}{*}{0.21}&\multirow{2}{*}{3.2}	 	\\ &&&&&&&&\\
			
			\multirow{2}{*}{$N_T  = 2$ }& 
			\multirow{2}{*}{2.2}&\multirow{2}{*}{0.57}	& 
			\multirow{2}{*}{3.8}&\multirow{2}{*}{0.45}	& 
			\multirow{2}{*}{0.44}&\multirow{2}{*}{9.9}	& 
			\multirow{2}{*}{0.23}&\multirow{2}{*}{7.9}	 	\\ &&&&&&&&\\
			
			\multirow{2}{*}{$N_T   \geq 3$ }& 
			\multirow{2}{*}{0.6}&\multirow{2}{*}{0.99}	& 
			\multirow{2}{*}{0.8}&\multirow{2}{*}{0.55}	& 
			\multirow{2}{*}{0.22}&\multirow{2}{*}{24.8}	& 
			\multirow{2}{*}{0.34}&\multirow{2}{*}{86.1}	 	\\ &&&&&&&&\\
			
			\hline
		\end{tabular}
		\caption{Segmentation of the entire phase-space based on the information of number of tracks associated to the jets and the HCAL information. The regions with $\theta < \theta_0$ are dominated by the ECAL energy deposition (less HCAL deposition) while $\theta \geq \theta_0$ 
is the same but with HCAL information only where we choose $\theta_0$ = 0.25. Each sample has been divided into two regions, one ECAL-rich while other HCAL-rich. We then further separate them in terms of number of charged tracks associated to the leading jet. For details, see the text.}   
		\label{tab:seg}
	\end{center}
\end{table}
\begin{table}[h!]
        \begin{center}
                \begin{tabular}{c||c|c|c|c}
\hline
                        \multirow{2}{*}{$\theta_J$} &  
                        \multirow{2}{*}{$N_{T} = 0$}&
                        \multirow{2}{*}{$N_{T} = 1$}&
                        \multirow{2}{*}{$N_{T} = 2$}&
                        \multirow{2}{*}{$N_{T} \geq 3$}         \\ &&&&\\
\hline
\hline
                        \multirow{2}{*}{$\theta_{J} < \theta_0$}  &
                        \multirow{2}{*}{EC0} &
                        \multirow{2}{*}{EC1}&
                        \multirow{2}{*}{EC2}&
                        \multirow{2}{*}{EC3+}              \\ &&&&\\
\hline
                        \multirow{2}{*}{$\theta_{J} > \theta_0$}  &
                        \multirow{2}{*}{HC0} &
                        \multirow{2}{*}{HC1}&
                        \multirow{2}{*}{HC2}&
                        \multirow{2}{*}{HC3+}              \\ &&&&\\
\hline
                \end{tabular}
                \caption{The nomenclature of the regions based on the charged track multiplicity and calorimetry information. }
                \label{tab:seg-names}
        \end{center}
\end{table}
In Table~\ref{tab:seg} we display the result of segmenting the entire phase-space based on $\theta_J$ and $N_T$.  For a jet of kind $\alpha$, we define the efficiency in a patch/bin as 
\begin{equation}
\epsilon_\alpha (\text{bin}) \ = \ \frac{\text{Number of jets of type }\alpha \text{ in the bin}}{ \text{Total number of jets of type } \alpha}
\end{equation}

Additionally, in Table~\ref{tab:seg-names}, we denote how we refer to these regions in this work.   
For example, the segment EC1, represents the region occupied by the jets with $\theta < \theta_0$ and $N_T =  1$, whereas the segment HC2 represents the region occupied by jets with $\theta \geq \theta_0$ and $N_T =  2$. As seen from the left plot in Fig.\ref{thetaJ&NT}, one expects regions with $\theta \geq \theta_0$ and a large number of tracks are rich in parton initiated jets and further binning these jets in $N_T$ does not really help in finding regions relatively free of QCD-jets. We simply group these regions occupied by HCAL rich jets with large tracks under the designation HC3+.

\subsection{\label{sec:4.2}A realistic veto using multivariate analyses}

Once we segment the entire phase-space in terms of number of tracks and energy profile associated to a jet of standard objects, next goal is to find regions of the phase-space where the contribution coming from these standard objects are at the sub-percent level.  
We incorporate all the variables discussed in Sec.\ref{sec:3}, important in terms of its discrimination power, and then perform a multivariate analysis in order to achieve the maximum sensitivity.  

As explained in the guideline discussed in Sec.\ref{sec:2.3}, we begin with constructing three BDTs, namely 
\begin{enumerate}
\item  $\Bgj :$ A BDT to separate photons (signal) from QCD-jets (background).  
\item  $\Bjt :$ A BDT to separate QCD-jets (signal) from taus (background).     
\item  $\Bgt :$  A BDT to separate photons (signal) from taus (background).   
\end{enumerate}

The working principle in a BDT is straightforward. It is a collection of decision trees whose main purposes are to pairwise discriminate two samples. For the sake of notation we refer to one sample as `signal' and the other as `background'.  Each tree is characterized by different levels of hard cuts on the variables, which selects regions rich in signals. Since a single tree can be sensitive to the choice of the cuts on the variables, multiple trees are constructed, which is followed by a weighing procedure. As mentioned before, the final outcome of the BDT is a single real number  (namely, the `response') for each object in the sample. We reweigh responses such that it lies in the  range $0$ to $+1$. For a good discriminator,  the background and signal events are characterized by $r \sim  0$ and $r \sim +1$ respectively. 

In our case, the samples consist of jets. In    $\Bgj $, for example, we call the set of photons (or $\{  J_\gamma\}$) as signals and  the set of QCD-jets (or $\{  J_j\}$) as  backgrounds.  Corresponding to a decision, each jet in the sample (mixed signal and background) is assigned a response of the given analysis. 
In this example, we expect responses for photons to lie at around $1$, whereas QCD-jets to accumulate around $0$. Further, as explained in Sec.\ref{sec:2.3}, we use a naming convention for the responses, similar to the BDTs. For example, the responses for $\Bgj$ will be denoted by $\rgj$. 

A crucial part for the construction of BDTs is to find a set of variables. Even though one can use the full set of variables described in Sec.\ref{sec:3.2.1} for all the BDTs,  we rather make judicious choices for each of the BDTs. For example,   for $\Bgj$, we select variables which exhibit good discriminatory power between photons and QCD-jets. In Table~\ref{discriminatingvariables}, we provide the list of the variables we consider for the three BDTs.   
\begin{table}[htb!]
\begin{center}
	\begin{tabular}{ |c|c| }
\hline
		\multirow{2}{*}{BDT} &   \multirow{2}{*}{Variables} \\  & \\
			  
\hline
		\multirow{2}{*}{$\Bgj$} & 
			\multirow{2}{*}{$\lambda_J,C_1,r_1$}  \\	& \\
							
		\multirow{2}{*}{$\Bjt$ } & 
			\multirow{2}{*}{$\lambda_J,r_2,\tau_{31} $} \\	& \\
				
		\multirow{2}{*}{$\Bgt$}  & 
			\multirow{2}{*}{$\epsilon_J,C_1,\lambda_J,r_1,e_2$} \\	& \\
\hline
	\end{tabular}
\end{center}
\caption{List of variables for the discrimination of a given pair of standard-jets. The first variable represent the one best suited (highest weighted) for this discrimination when $\theta_J$ and $N_T$ are excluded.}
\label{discriminatingvariables}
\end{table}
\begin{figure}[htb!]
        \begin{center}
         	\includegraphics[width=\textwidth]{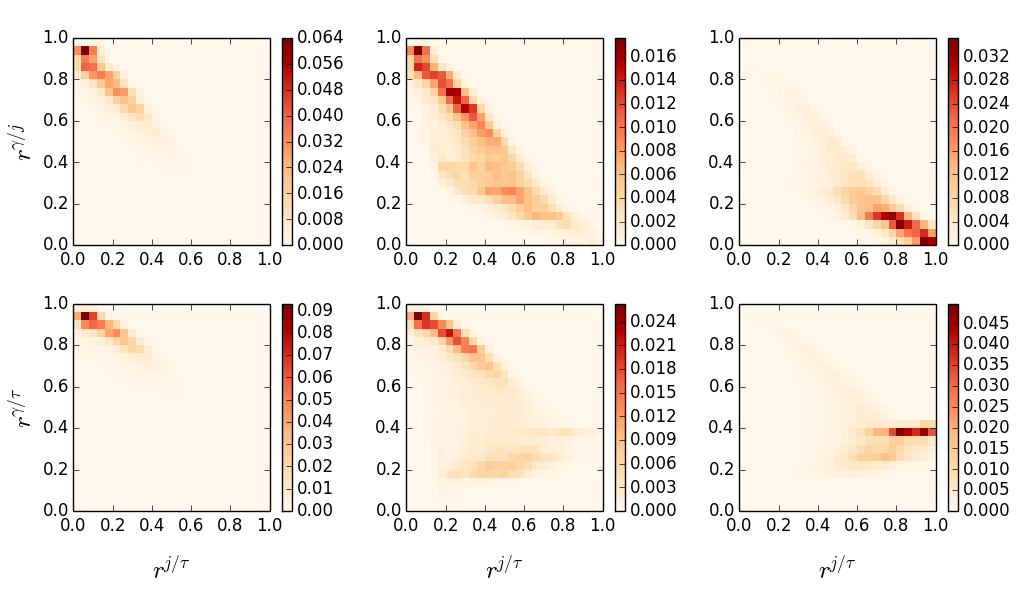} \\
        \end{center}
        \caption{ \label{fig:Responses_2d} The 2-dimensional distributions of the BDT response variables $\rgj$,  $\rjt$, and $\rgt$ for the standard-jets. The columns from the left to right represent the distributions for the photon, $\tau$ and QCD-jets respectively. In these plots we have used $2D$-bins of size $\left(0.04\times 0.04\right)$ in units of responses.}
\end{figure}

We summarize the results of the BDT analyses (responses) in Fig.\ref{fig:Responses_2d}.  Each of the plots in Fig.\ref{fig:Responses_2d} shows two dimensional probability distributions corresponding to various standard-jets. The left column corresponds to responses for photons: the top plot shows $2D$-histogram in $\rjt$--$\rgj$ plane, whereas the bottom plot shows that in $\rjt$--$\rgt$ plane. The color coding associated with each bin reflects the  \emph{probability} (not probability-density) of a photon to occupy the bin.  The physics understanding of these plots are simple. Note that the $y$ axes in both the plots represent responses for the BDTs $\Bgj$ and $\Bgt$, which treat photons as signals and therefore assign large responses correctly.  As far as the $x$-axis is concerned, the BDT $\Bjt$ considers photons more $\tau$-like (background) than qcd-jets (signal). Therefore, photons show up mostly in top left corner in both the figures. 

The central column of plots in  Fig.\ref{fig:Responses_2d} show the same distributions, but for $\tau$s. These follow patterns quite similar to that of the photons, and occupy mostly in the top left corner of both the plots. A striking feature in both the plots is that there is quite a few of these jets get characterized by large responses under BDT $\Bjt$ even though $\tau$s are treated as background jets. This suggests that the characteristics identified by $\Bjt$ to separate $j$ from $\tau$, does not perform as well for a small fraction of tau jets. We think that $\Bjt$ becomes efficient in separating taus with single prongs (the largest fraction of tau samples) from QCD-jets.  In fact support for this argument can be found in the $\Bjt$ responses for photons, which assign all photons  (single pronged)  small responses. Taus with multi-prong structures show us with large responses.  The response of $\Bgt$, on the other hand, is quite disappointing. It simply shows that the variables we select here, which mostly analyzes the transverse features of energy depositions in the calorimeter cells are not very efficient in discriminating photons from the most of the tau samples (mostly single pronged). The substructure variables only manage to find taus with multi-prong structures to be substantially different from the photon samples. 

Finally, the rightmost column in  Fig.\ref{fig:Responses_2d} we show the same probability distributions for QCD-jets. The top plot does not require any subtle explanation. The BDTs $\Bjt$ and $\Bgj$ treat qcd-jets as signals and backgrounds respectively, giving these preferred positions in the bottom right corner. The bottom plot is quite interesting. The BDT, $\Bgt$ is  trained on discriminating photons from the taus. Even though it does not turn out to be very good at separating taus from photons, it nevertheless assigns most of qcd-jets responses within a narrow zone. As we show later, it will end up being highly useful in constructing a veto for QCD-jets.

One can use the phase space distributions to construct vetoes for these standard objects. For example, the region rich in QCD-jets can be roughly parameterized as:
\begin{equation}
	 C_1 \leq \rgt  \leq C_2    \qquad  \text{AND}  \qquad \Bigl( \
	 \left| \rgj + \rjt -1.0 \right|  \leq C_3 \qquad  \text{OR}   \qquad
	  \left| \rgj - \rjt  \right|  \geq C_4 \ \Bigr)  \, . 
\label{eq:vetobar_j}
\end{equation}
In the above equation, $C_i$s are parameters that can be adjusted to contain most of QCD-jets. A QCD-veto will then reject all jets in the phase-space described in Eq.~\eqref{eq:vetobar_j}. 

In this work, instead of finding a region rich in QCD-jets by eye, we rather take a different approach in order to construct a QCD-veto. We discretize the 3$D$ space of responses $\{ \rjt, \rgj, \rgt \}$ into bins; we calculate the probability of finding QCD-jets in each of the bins; we sort bins in decreasing probability; and finally keep vetoing sorted-bins until only a small (desired) fraction of QCD-jets remain.  

Let us elaborate on the procedure described above with a concrete example. Consider the region HC2. As reflected in Table~\ref{tab:seg},  in HC2 $\epsilon_j = 0.079$. This implies that  $7.9\%$ of all QCD-jets occupy this section of  the phase space. The goal of the following exercise will be to reduce QCD rate below an acceptable level, say $R_j$. In short we want $\epsilon_j \leq R_j$ in the region HC2.  
\begin{itemize}
\item  We begin with binning the full phase space into cubes of sizes $\left(0.04\times 0.04 \times 0.04\right)$ in units of responses. We can represent each bin either in $3D$ (for example, the bin $(i,j,k)$  represents the $i$-th in $\rjt$ direction, $j$-th in $\rgj$ direction, and $k$-th in $\rgt$ direction), or in $1D$ (for example, the $(i,j,k)$-th bin gets represented as the $b$-th bin, where $b = i + n_b \times j + n_b^2 \times k$ with $n_b$ being the number of bins, here $n_{b} = 25$.). 

\item Each bin is characterized by the probability of QCD-jets occupying the bin. In particular we define bin probabilities to be 
\begin{equation}
P_b \ = \ \frac{1}{N} \sum_j \left\{ \quad
\begin{aligned}
1 & \qquad  \text{if} \ j \in b \\
0 & \qquad  \text{if} \ j \notin b 
\end{aligned}
\right. \quad  ,
\end{equation}
where $N$ is the total number of QCD-jets studied and the index $j$ runs over all QCD-jets.  Also, clearly by construction $\sum_b P_b = 1$.  The cumulative probability of each bin (namely, $C_b$) is defined as 
\begin{equation}
C_b \ =  \sum_{b'} \left\{ \quad
\begin{aligned}
P_{b'} & \qquad  \text{if} \ P_{b'}  \geq P_b \\
0 & \qquad  \text{else} 
\end{aligned}
\right. \quad ,
\end{equation}
where we sum over all bins $b'$.  A better pictorial representation can be obtained if bins are sorted in decreasing probabilities as shown in Fig.\ref{fig:SortedResponse2D_QCD}. In the left-most plot we have shown the distributions $P_b$ and $C_b$ for QCD jets by solid and dashed lines respectively. Note that $C_b$ asymptotes towards $1$ as per expectations.

\item We also determine bin probabilities in each segment. For example the bin probabilities in HC2 will be given by 
\begin{align}
\label{eq:segP}
P_b^\text{HC2} \ & = \ \frac{1}{N} \sum_j \left\{ \quad
\begin{aligned}
1 & \qquad  \text{if} \ j \in b \quad \& \quad \theta \geq \theta_0 \quad \& \quad N_T = 2  \\
0 & \qquad  \text{else}
\end{aligned}
\right. \quad . \\
C_b^\text{HC2} \ & =  \sum_{b'} \left\{ \quad
\begin{aligned}
P_{b'}^\text{HC2} & \qquad  \text{if} \ P_{b'}^\text{HC2}  \geq P_b^\text{HC2} \\
0 & \qquad  \text{else} 
\end{aligned}
\right. \quad .
\end{align}
Note that the denominator in Eq.~\eqref{eq:segP} is still given by the total number of QCD-jets. Therefore, one gets $\sum_b P_b^{\text{HC2}} = \epsilon_j^{\text{HC2}}$.  In the central plot of Fig.\ref{fig:SortedResponse2D_QCD} we 
show $P_b^\text{HC2}$ and $C_b^\text{HC2}$ again by solid and dashed lines respectively. The distribution  
$C_b^\text{HC2}$ now asymptotes to $ \epsilon_j^{\text{HC2}}$.

\begin{figure}[htb!]
	\includegraphics[width=\textwidth]{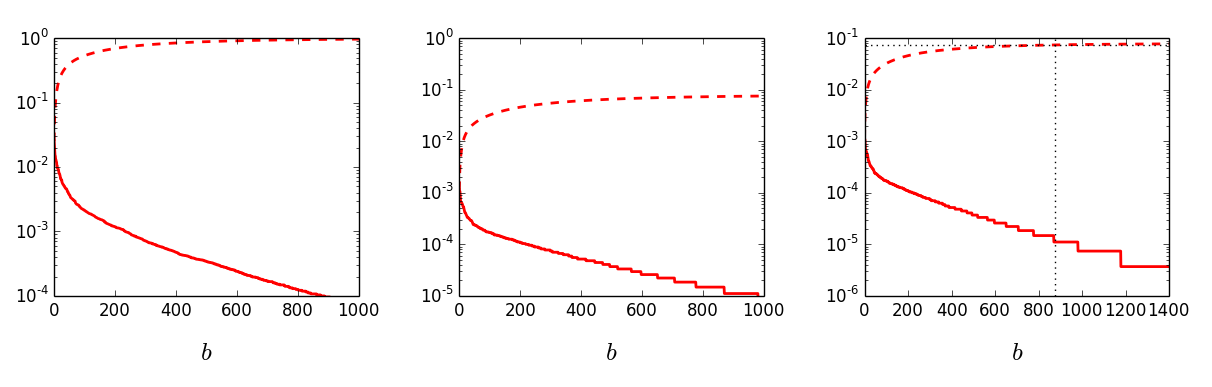} 
	\caption{\label{fig:SortedResponse2D_QCD} Left: The distributions $P_b$ (solid) and $C_b$ (dashed) for QCD-jets. The bins are sorted in decreasing $P_b$ and we only show the first $1000$ bins. Center: The distributions  $P_b^{\text{HC2}}$ (solid) and $C_b^{\text{HC2}}$ (dashed). Right: The same plot as the center one, but for the first $1500$ bins. The dotted horizontal line represents $y=0.074$ (see text for explanation). The QCD-veto for HC2 as described in Eq.~\eqref{eq:qcd-veto} and in Eq.~\eqref{eq:qcd-veto2} blocks $875$ bins left of the right dotted line.}
\end{figure}

\item The \emph{QCD-veto} is simply about blocking a collection of bins rich in QCD-jets  so that only a small fraction of QCD-jets are allowed.  Given a tolerance rate $R_j$ (defined as the rate at which QCD-jets can be allowed), one can then determine the QCD veto function (for HC2) using
\begin{equation}
f_b^\text{j} (\text{HC2}) \ = \  \left\{ \quad
\begin{aligned}
1 & \qquad  \text{if} \ C_{b}^\text{HC2}  \geq \left( \epsilon_j^{\text{HC2}} - R_j \right) \\
0 & \qquad  \text{else} 
\end{aligned} 
\right.
\label{eq:qcd-veto} 
\end{equation}
where $0$ represents bins vetoed and $1$ the bins accepted. The logic behind the equation above can be explained in the rightmost plot in Fig.\ref{fig:SortedResponse2D_QCD}. The plot is identical to the middle plot except that we only plot first $1500$ bins. The dotted horizontal line represents at $y  = \epsilon_j^{\text{HC2}} - R_j = 0.074$ (here we have taken $R_j = 0.005$ and $\epsilon_j^{\text{HC2}}$ is given as $0.079$ from Table~\ref{tab:seg}). The vertical dotted line represents the bin for which $C_{b}^\text{HC2}  = \left( \epsilon_j^{\text{HC2}} - R_j \right) = 0.074 $. The veto function in Eq.~\eqref{eq:qcd-veto} simply vetoes bins on the left of the line and accepts the bins on the right. The veto function in Eq.~\eqref{eq:qcd-veto} can be rewritten in terms of $P_b$ as well. Naming  the point where the vertical line intersects $P_{b}^\text{HC2} $ to be $P_{R_j}(\text{HC2})$, we can restate
\begin{equation}
f_b^\text{j} (\text{HC2}) \ = \  \left\{ \quad
\begin{aligned}
1 & \qquad  \text{if} \ P_{b}^\text{HC2}  \leq  P_{R_j}(\text{HC2}) \\
0 & \qquad  \text{else}. 
\end{aligned} 
\right.
\label{eq:qcd-veto2} 
\end{equation}
Note that vetoes as stated in Eq.~\eqref{eq:qcd-veto} and in Eq.~\eqref{eq:qcd-veto2} are slightly different, may yield slightly different values of $\epsilon_j$ after vetoes are enforced. Differences arise since we did not impose strict inequalities (rather we use $\geq$ and $\leq$), which get magnified especially in case there are multiple bins corresponding to the $P_{b}^\text{HC2}  =  P_{R_j}(\text{HC2})$.       
\end{itemize}

\begin{table}[htb!]
        \begin{center}
                \begin{tabular}{|c||c||c|c|c|c|}
\hline
                        \multirow{2}{*}{Regions}  & \multirow{2}{*}{Vetoes used} &
                        \multirow{2}{*}{$\epsilon_\gamma$ (in $\%$)}&
                        \multirow{2}{*}{$\epsilon_e$ (in $\%$)}&
                        \multirow{2}{*}{$\epsilon_\tau$ (in $\%$)}&
                        \multirow{2}{*}{$\epsilon_j$ (in $\%$)}         \\ &&&&&\\
\hline

                        \multirow{2}{*}{EC0}  &
                        \multirow{2}{*}{Photon Veto with $R_{\gamma} = 0.05$} &
                        \multirow{2}{*}{5.0}&
                        \multirow{2}{*}{0.70}&
                        \multirow{2}{*}{0.59}&
                        \multirow{2}{*}{0.07}              \\ &&&&&\\

                        \multirow{2}{*}{EC1}  &
                        \multirow{2}{*}{Electron Veto with $R_{e} = 0.05$} &
                        \multirow{2}{*}{0.93}&
                        \multirow{2}{*}{5.0}&
                        \multirow{2}{*}{3.8}&
                        \multirow{2}{*}{0.13}              \\ &&&&&\\

                        \multirow{2}{*}{EC2}  &
                        \multirow{2}{*}{No Veto} &
                        \multirow{2}{*}{2.1}&
                        \multirow{2}{*}{3.8}&
                        \multirow{2}{*}{0.44}&
                        \multirow{2}{*}{0.22}              \\ &&&&&\\

                        \multirow{2}{*}{EC3+}  &
                        \multirow{2}{*}{No Veto} &
                        \multirow{2}{*}{0.61}&
                        \multirow{2}{*}{0.81}&
                        \multirow{2}{*}{0.22}&
                        \multirow{2}{*}{0.34}              \\ &&&&&\\

\hline \hline

                       \multirow{2}{*}{HC0}  &
                        \multirow{2}{*}{QCD Veto with $R_j = 0.005$} &
                        \multirow{2}{*}{2.4}&
                        \multirow{2}{*}{0.34}&
                        \multirow{2}{*}{3.1}&
                        \multirow{2}{*}{0.58}              \\ &&&&&\\

                        \multirow{3}{*}{HC1}  &
                        \multirow{2}{*}{QCD Veto with $R_j = 0.005$} &
                        \multirow{3}{*}{0.21}&
                        \multirow{3}{*}{0.66}&
                        \multirow{3}{*}{4.9}&
                        \multirow{3}{*}{0.25}         \\    
                   
                         & \multirow{2}{*}{Tau Veto with $R_{\tau} = 0.05$} &&&&\\	 
                         &&&&&\\

                        \multirow{2}{*}{HC2}  &
                        \multirow{2}{*}{QCD Veto with $R_j = 0.005$} &
                        \multirow{2}{*}{0.14}&
                        \multirow{2}{*}{0.08}&
                        \multirow{2}{*}{2.3}&
                        \multirow{2}{*}{0.55}              \\ &&&&&\\

                        \multirow{2}{*}{HC3}  &
                        \multirow{2}{*}{QCD Veto with $R_j = 0.005$} &
                        \multirow{2}{*}{0.06}&
                        \multirow{2}{*}{0.03}&
                        \multirow{2}{*}{5.2}&
                        \multirow{2}{*}{0.54}              \\ &&&&&\\

                        \multirow{2}{*}{HC4+}  &
                        \multirow{2}{*}{QCD Veto with $R_j = 0.005$} &
                        \multirow{2}{*}{0.04}&
                        \multirow{2}{*}{0.02}&
                        \multirow{2}{*}{0.21}&
                        \multirow{2}{*}{0.56}              \\ &&&&&\\

\hline
                \end{tabular}
                \caption{Efficiencies in various segments after vetoes are imposed. The vetoes are applied in a way so that QCD-jets are allowed only at the level of $\lesssim 0.5\%$, whereas for other standard-jets we allow efficiencies of order $\lesssim 5\%$.}
       	\label{table:eff_vetoes}
        \end{center}
\end{table}

We impose QCD-veto as described above in all HC segments.  Similar constructions are used to construct photon-veto (for EC0), electron-veto (for EC1 and EC2), and tau-veto (for HC1). The procedure is identical except that we can allow for a larger rate for other vetoes.  To be specific, we mainly use two different target rates
\begin{equation}
R_j \ = \ 0.005 \, , \qquad \text{and} \qquad R_\gamma \ = \ R_e \ = \ R_\tau \ = \ 0.05 \, . 
\end{equation}
This implies that we target blocking order $199$ in $200$ (or target allowing only $1$ in every $200$)  QCD-jets. For jets of other types, we could be less restrictive and allow more jets to pass through (since the production rate for these jets are small compared to the QCD-jets). In particular, we try blocking roughly $19$ out of $20$ photons, for example.  Note that these number are in sync with what we typically target as tolerable mis-tagging efficiency when designing a tagger. For example, in standard jet-flavor-tagging procedure the working point typically involves $1\%$ or higher mistag efficiency from light-flavor QCD-jets. Similarly, for photon tagging, we tolerate around $5$-$6\%$ of mistag from electrons. 

In Table~\ref{table:eff_vetoes} we show the results as we impose vetoes judiciously on different segments.  It turns out that single vetoes are efficient enough to bring down the rate of standard-jets below the acceptable range in all but one segment. In HC2, we need a tau-veto along with a QCD-veto. Note that, given our target, we do not need any veto for EC2, and EC3+, since these segments are already pure. 

\section{\label{sec:5}Example Non-standard objects after vetoes}

The generality of our analysis enables its application across a wide range of models which includes various non-standard objects, e.g., highly collimated particles, long lived particles etc.  In this section, we discuss, as an example, the sensitivity of this analysis to capture some of these non-standard objects, especially collimated di-photon, di-electron and di-tau samples. 

Let us emphasize that the purpose of this section is not to categorize, describe or even to tabulate all possible  anomalous objects -- simply because such tasks are more or less rendered less important due to the nature of our proposal. The vetoes are constructed around the standard objects only, and thus we can always be agnostic of the exact form of new physics while attempting to find traces of new physics.  

In order to demonstrate the efficacy of our method, we take three examples of anomalous objects: 
\begin{enumerate}[i.]
\item Jets initiated by a pair of collimated photons. 
\item Jets initiated by a pair of collimated electrons.
\item Jets initiated by a pair of collimated taus (hadronic).
\end{enumerate}
Note again that vetoes we use (as tabulated in Table~\ref{table:eff_vetoes}), have no information regarding the exact nature of any of these anomalous objects. Of course, if analyses use this information, they would perform better -- the job here is to demonstrate that even without using any information of anomalous objects we can capture decent amount of these objects. 

In order to evaluate the rate at which these objects pass the vetoes, we first need to generate samples, which requires a toy Lagrangian. Once again the details of Lagrangian does not matter.  Following the example shown in Ref.~\cite{Ellis:2012zp,Ellis:2012sd}, we consider a handful of toy models here.  The simplistic model by extending the SM Higgs sector with a new scalar field (say, $n_1$) can be written as:  
\begin{equation}
\mathcal{L}_\text{toy1} \ =  \  \frac{1}{2}\left( \partial^2 - m_{1}^2 \right)n_1^2 \ + \  
		\frac{1}{2} \:\mu_{1} h n_1^2 
\ + \ \eta_{a} \frac{1} {\Lambda} \: n_1 F^{\mu\nu} \tilde{F}_{\mu\nu} \ +\   \eta_{e} \: n_1 e e^c 
\ + \ \eta_{\tau}  \: n_1 \tau \tau^c \, ,
\label{eq:toy_lagrangian1}
\end{equation}
where $h$ represents the SM Higgs scalar (of mass $m_h \sim 125\gev$);  $m_1, \mu_1$ are masses much smaller than the cut-off $\Lambda$; and finally all $\eta_i$ are dimensionless constants.  Now, the limit $\eta_e, \eta_\tau \rightarrow 0$, gives rise to Higgs decay to four photons via $p~p\rightarrow h \rightarrow n_1(\gamma \gamma) n_1(\gamma \gamma)$. In the limit, $m_{1} \ll m_h$, one actually finds each $n_1$ giving rise to a collimated pair of photons (say, the diphoton-jets). Similarly in the same limit, one finds dielectron-jets or ditau-jets for $\eta_a, \eta_\tau \rightarrow 0$ or  $\eta_a, \eta_e \rightarrow 0$ respectively.

We further emphasize that  we only use this to generate sample of anomalous objects that tests our proposed anomaly finder. While the Lagrangian is Eq.~\eqref{eq:toy_lagrangian1} is easy to understand as well as to implement in a Monte Carlo, the use of Higgs scalar always raises the question whether we can search of it indirectly just using some variations of current search strategies. Such questions are irrelevant. If Higgs is replaced by a new particle of mass say, $1\tev$, which decays only to di-tau-jets, of course, no current strategy will work satisfactorily unless one devises a method to look for di-tau-jets in particular.

Note that the toy model in Eq.~\eqref{eq:toy_lagrangian1}, can be easily UV-completed in a electroweak symmetric model, where $n_1$ arise from a electroweak singlet. The mixing term with the Higgs scalar can arise from mixed quartic $\left| H \right|^2 n_1^2$, where $H$ is the electroweak doublet. This term also give rise to a quadratic piece in $n_1$, that gets absorbed in $m_1$. The term with electromagnetic gauge fields easily goes through with the replacement of $F_{\mu \nu} \rightarrow B_{\mu \nu}$, the field strength for hypercharge. Finally, terms with fermions break electroweak symmetry, and therefore must be proportional to the Higgs vacuum expectation value (namely, $v$). These terms, therefore, can arise from Higher dimensional terms (for example, $\frac{1}{\Lambda} n_1 H l_1 e^c$), where $l_1$ is the lepton electroweak doublet of the first generation.  The coupling $\eta_e$ is $v/\Lambda$ suppressed. 

The toy model can be extended easily to find non-standard jets with varied particle contents and topologies. A simple modification by adding a new scalar particle $n_2$, 
\begin{equation}
\mathcal{L}_\text{toy2} \ =  \  \mathcal{L}_\text{toy1} + \ 
	\frac{1}{2}\left( \partial^2 - m_{2}^2 \right)n_2^2 \ + \ 
		\frac{1}{2} \mu_{2} h n_2^2 +  \frac{1}{2} \mu_{12} n_1 n_2^2 .
\label{toy_lagrangian2}
\end{equation}
Now, setting $\mu_1$ to be zero in $\mathcal{L}_\text{toy2}$, opens up Higgs width to eight particles. Of course, in our preferred limit (\text{i.e.}, $m_{2} \ll m_h $), Higgs decays to two non-standard jets, with each of these standard-jets containing various combinations of four collimated particles. Exploring all sorts of topologies for a varied range of parameters is beyond the scope of this paper. As an example, we consider the Lagrangian in Eq.~\eqref{eq:toy_lagrangian1}, \textit{i.e.}, only study non-standard-jets consisting of pairs of photons, electrons, and tau particles. For the generation of the non-standard topologies, the parameters $\mu_1$ the decay of Higgs into two scalars ($n_1$) and is chosen to be $0.5$. The light scalar $n_1$ (of mass $m_{n_1} \sim 10\gev$) couples to a pair of photons, electrons and taus. To generate a collimated process, we assume the decay mode of $n_1$ into a given final state to be $100\%$. For instance, for the collimated photon topology, we assume $\eta_\gamma=1$ and set $\eta_e=\eta_\gamma=0$.
It is imperative to note that the decay of the Higgs ($h$) to a pair of $n_1$ with mass around 10$\gev$ provides the sufficient boost to $n_1$ (and thus to its decay products) so that it get clustered inside a single jet. 

Before proceeding,  we outline the behavior of the selected anomalous objects under the variables discussed in Sec.~\ref{sec:3}.
\begin{itemize}
\item \underline{$\log(\theta_J)$}: The left plot of Fig.\ref{sig:track} displays the distribution of the hadronic energy fraction in the leading jet for the non-standard jets. The di-photon (purple-dashed) and di-electron (blue-dotted) exhibit a behavior similar to the single photon and single electron jets as majority of both of the di-samples get deposited at the ECAL with no (or small) energy deposition at the HCAL. 
The di-tau jets, on the other hand, with both the taus decaying hadronically deposit a significant fraction of their energy in the HCAL, and thereby display a behavior similar to single $\tau$ and QCD jets. Thus, as expected, $\theta_J$ can be used efficiently to separate the ECAL-rich and HCAL-rich non-standard objects. 
\begin{figure}[!htb]
        \begin{center}
                \begin{tabular}{cc}
                        \includegraphics[width=7.2cm]{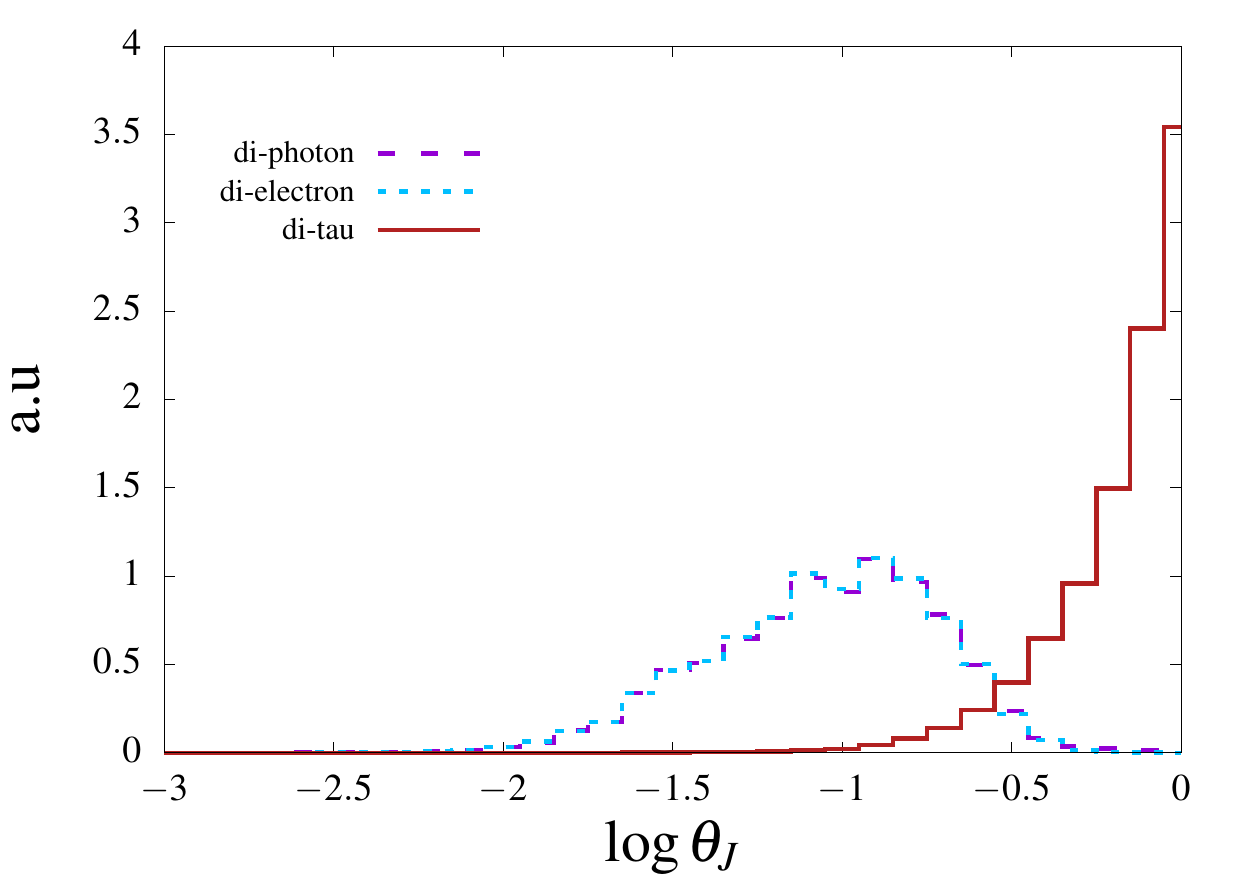}&
                        \includegraphics[width=7.2cm]{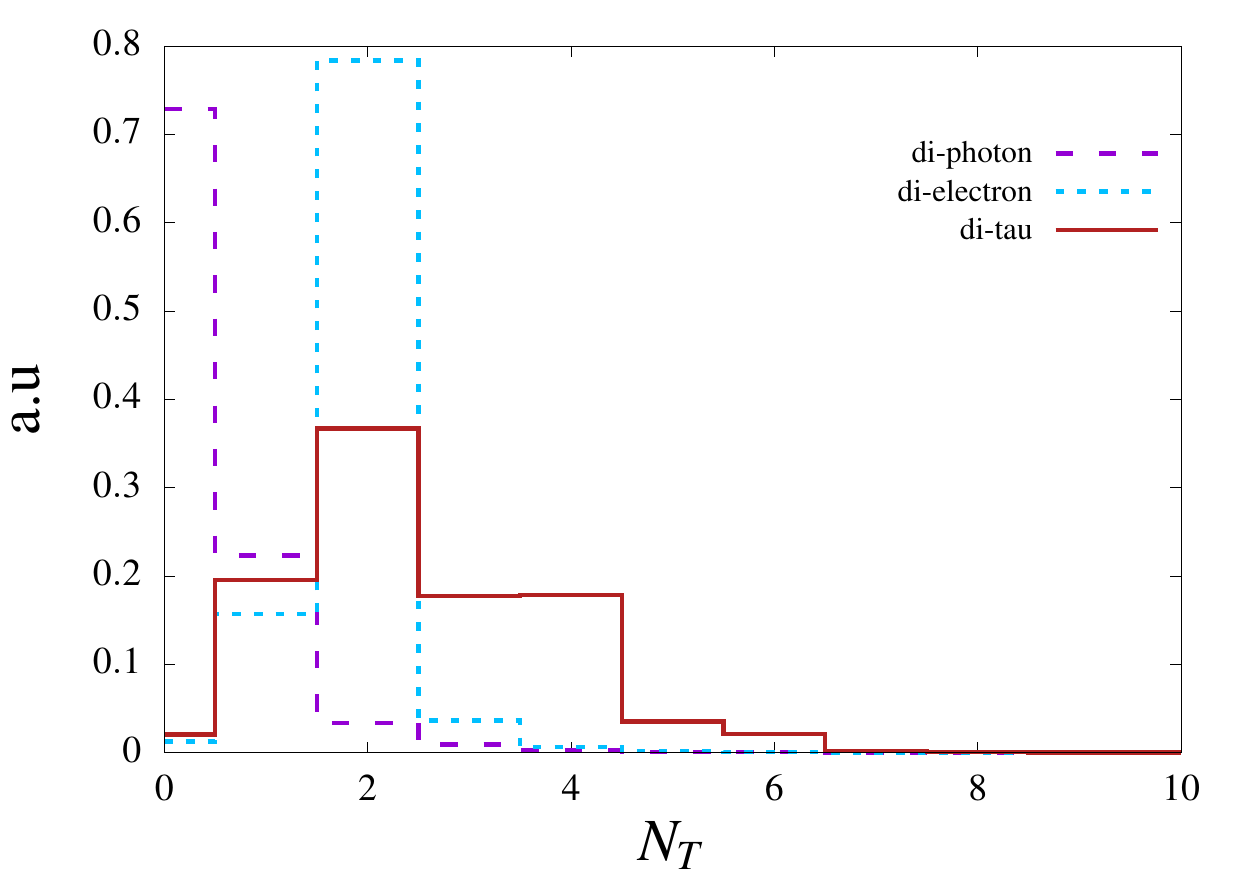}
                \end{tabular}
        \end{center}
        \caption{ The distribution of hadronic energy fraction (left) and the
number of tracks (right) in the leading jet for the non-standard objects.}
        \label{sig:track}
\end{figure}
\item \underline{$N_T$}: In the right plot of Fig.\ref{sig:track} we provide the distribution of the number of tracks inside the leading jet. The track multiplicity for the di-photon and the di-electron are expected to peak at 0 and 2 respectively, while for the di-tau it is a bit more involved owning the single or three pronged nature of a single tau (see Fig.\ref{thetaJ&NT}). As we observe the single-tau being dominantly single pronged, the corresponding track distribution for di-tau peaks at 2. However events with higher track multiplicities can be attributed to different combinations of the single and three pronged nature of the two taus inside the jet. Comparing Fig.\ref{thetaJ&NT} and Fig.\ref{sig:track} one can observe the track multiplicity distribution for the di-tau sample lies somewhat in between the single-tau (and other two di-samples) and QCD-jets, and thus $N_T$ (along with $\log\theta_J$) plays an important role while segmenting the phase space.  
\begin{figure}[htb!]
        \begin{center}
                \begin{tabular}{cc}
                        \includegraphics[width=7.2cm]{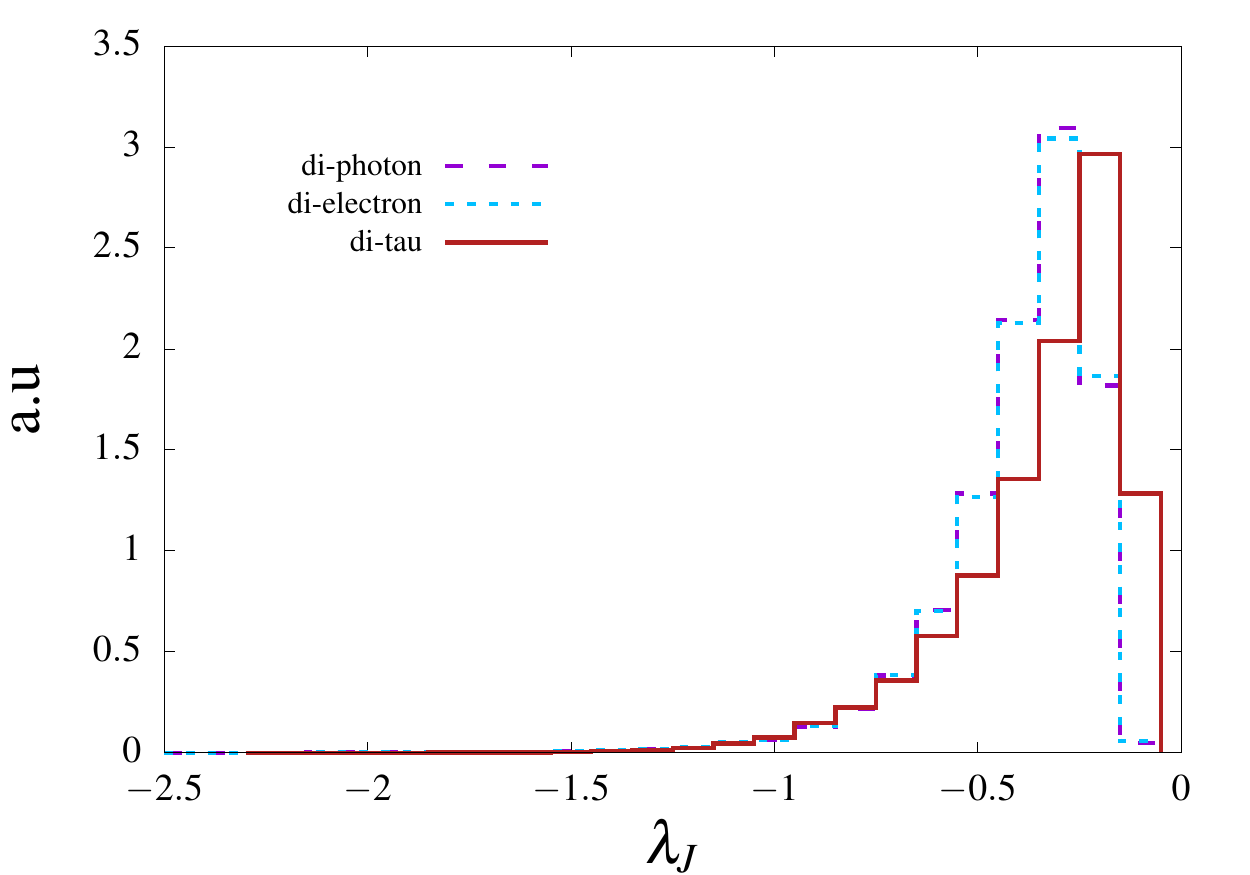}&\includegraphics[width=7.2cm]{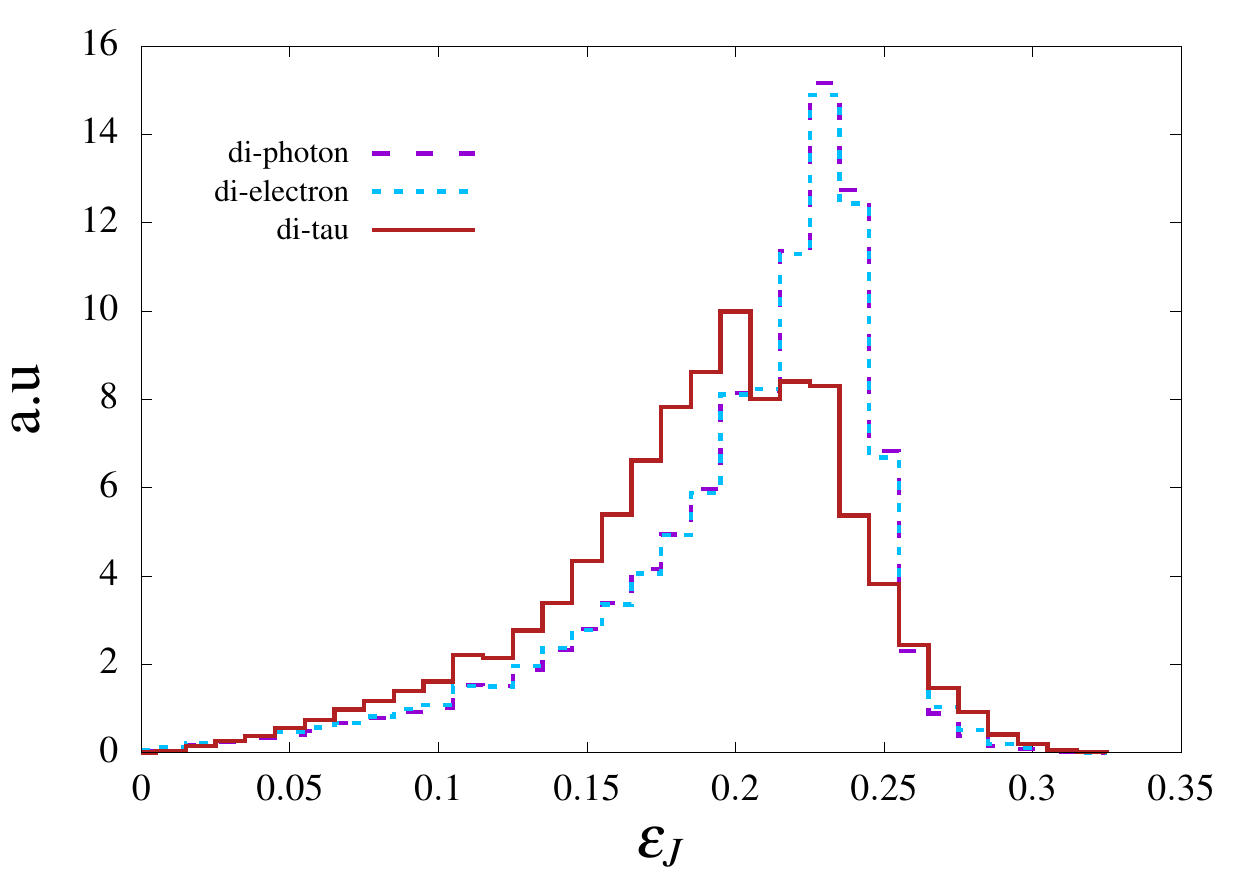}
                \end{tabular}
        \end{center}
        \caption{The distribution of $\lambda_J$ (left) and $\epsilon_J$ (right) for the non-standard objects.}
        \label{sig:lambda}
\end{figure}
\item \underline{$\lambda_J$ and $\epsilon_J$}: We have already discussed in Sec.\ref{sec:3}, $\lambda_J$ quantifies the fraction of the $p_T$ of the jet carried by the leading subjet. For single prong jets (with pencil like structure) $\lambda_J$ is expected to be small. For example, a jet with $\lambda_{J} > -0.3$  confirms the presence of two or more subjets. By construction the non-standard samples under consideration are of two prongs structure, as a result the distribution of $\lambda_J$ expectedly peaks at smaller negative values as opposed to the single electron, photon or tau jets, see Fig.\ref{sig:lambda}. From the distributions of $\lambda_J$, it is evident that the QCD-jets have a significant overlap with these non-standard objects. The behavior of $\epsilon_J$,  which is also a measure of the energy distribution inside a jet, exhibits a pattern similar to $\lambda_J$, see right plot of Fig.\ref{sig:lambda}. In this case also the non-standard jets have a pattern very much similar to the QCD-jets.
\begin{figure}[htb!]
        \begin{center}
                \begin{tabular}{cc}
                        \includegraphics[width=7.2cm]{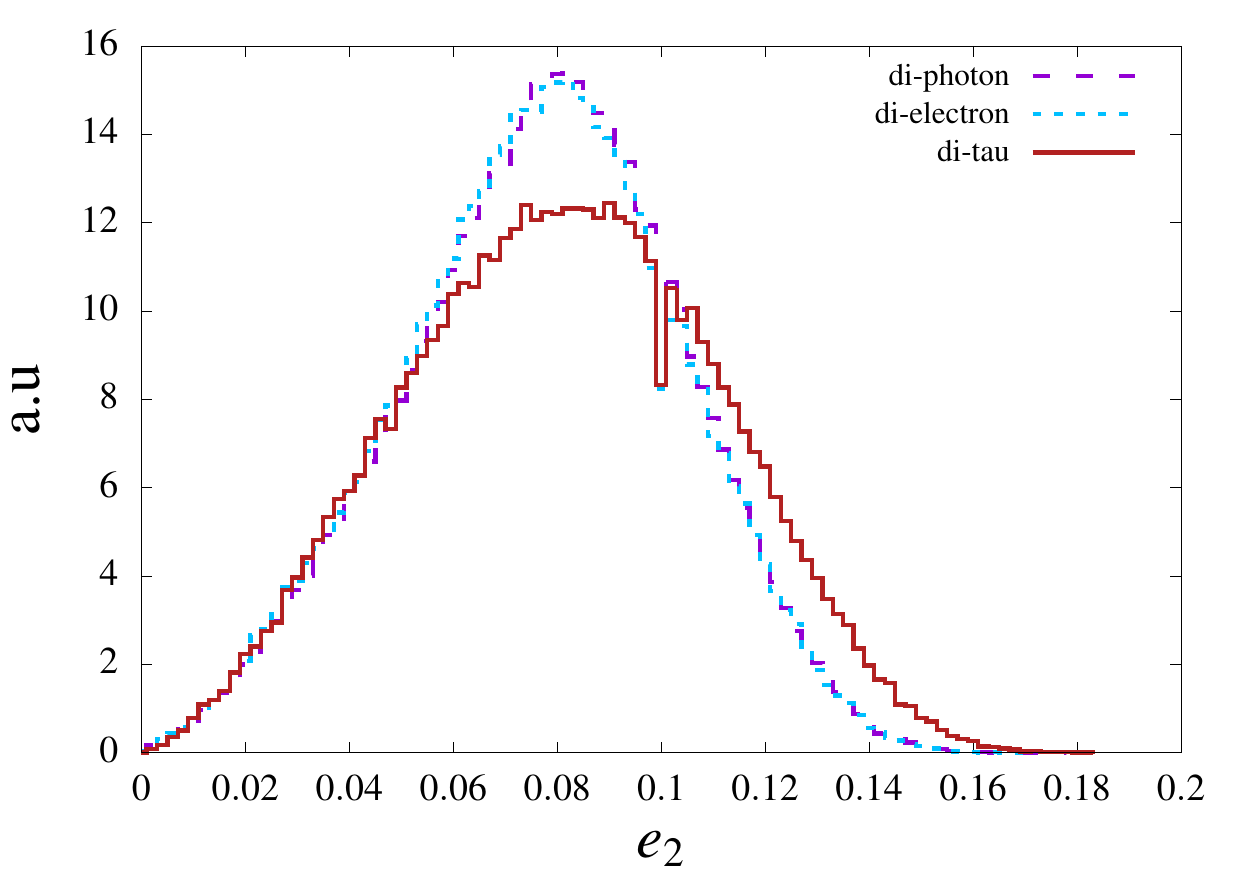}&\includegraphics[width=7.2cm]{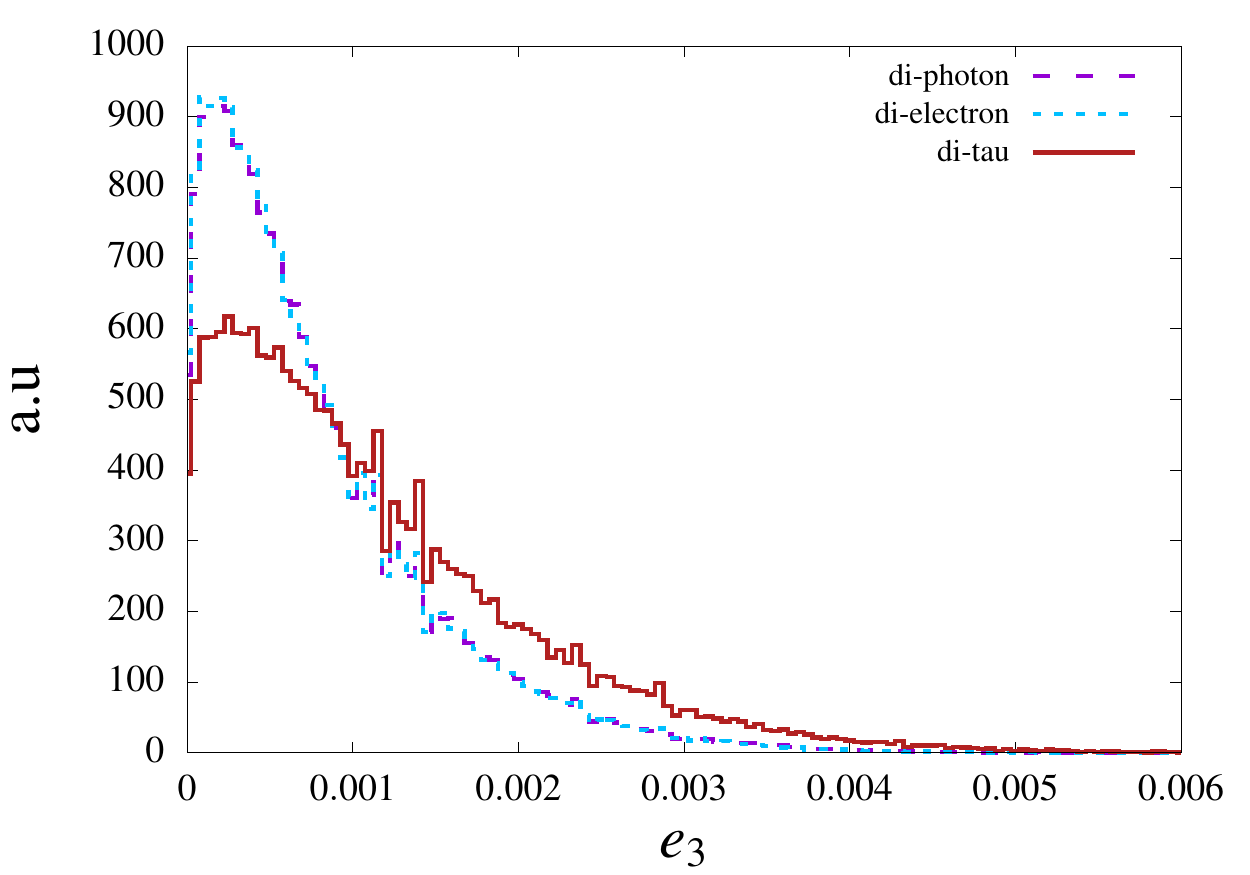}\\

                \end{tabular}
        \end{center}
        \caption{Distribution of the Energy Correlation Function (ECF) variables $e_2$ (left) and $e_3$ (right) for all the non-standard objects. }
        \label{sig:ecf1}
\end{figure}
\begin{figure}[htb!]
        \begin{center}
                \begin{tabular}{cc}
                        \includegraphics[width=7.2cm]{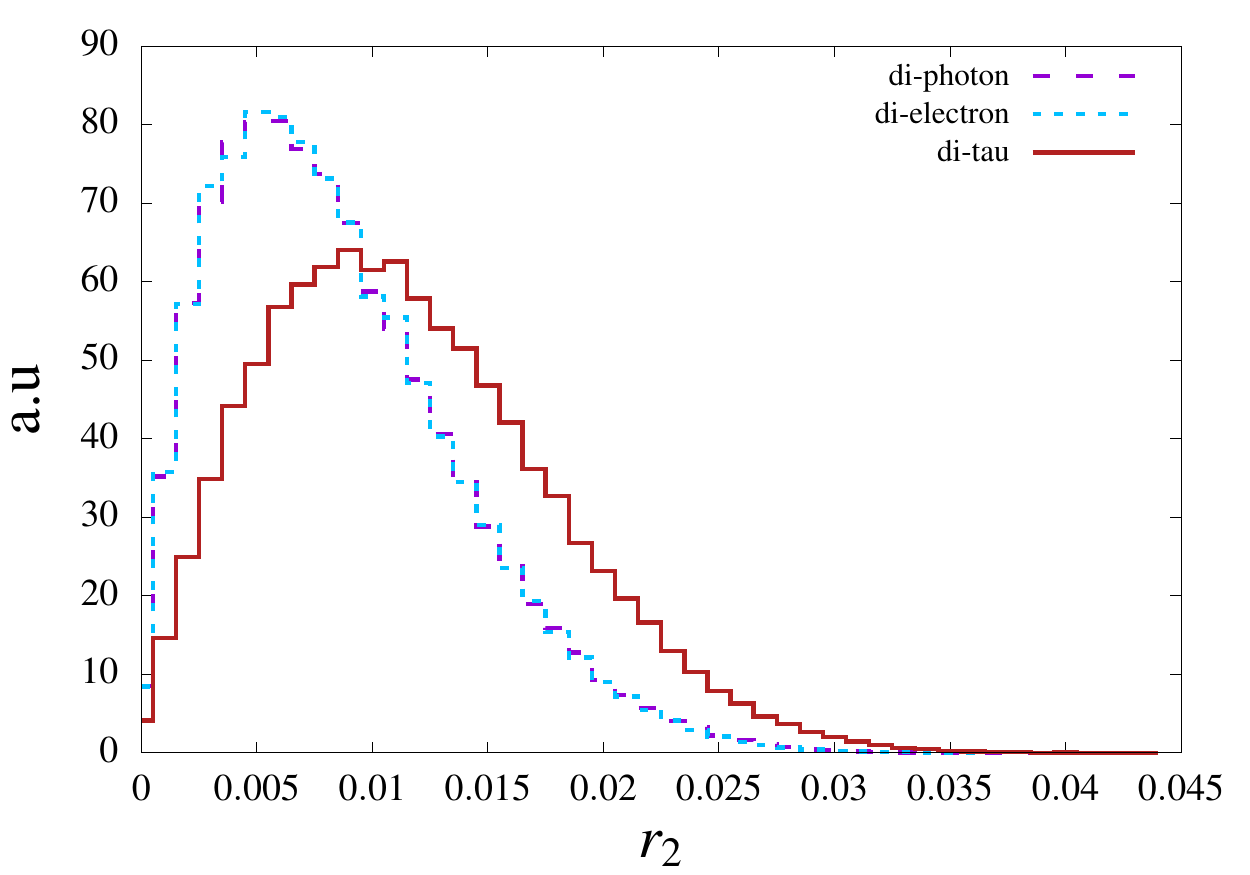}&\includegraphics[width=7.2cm]{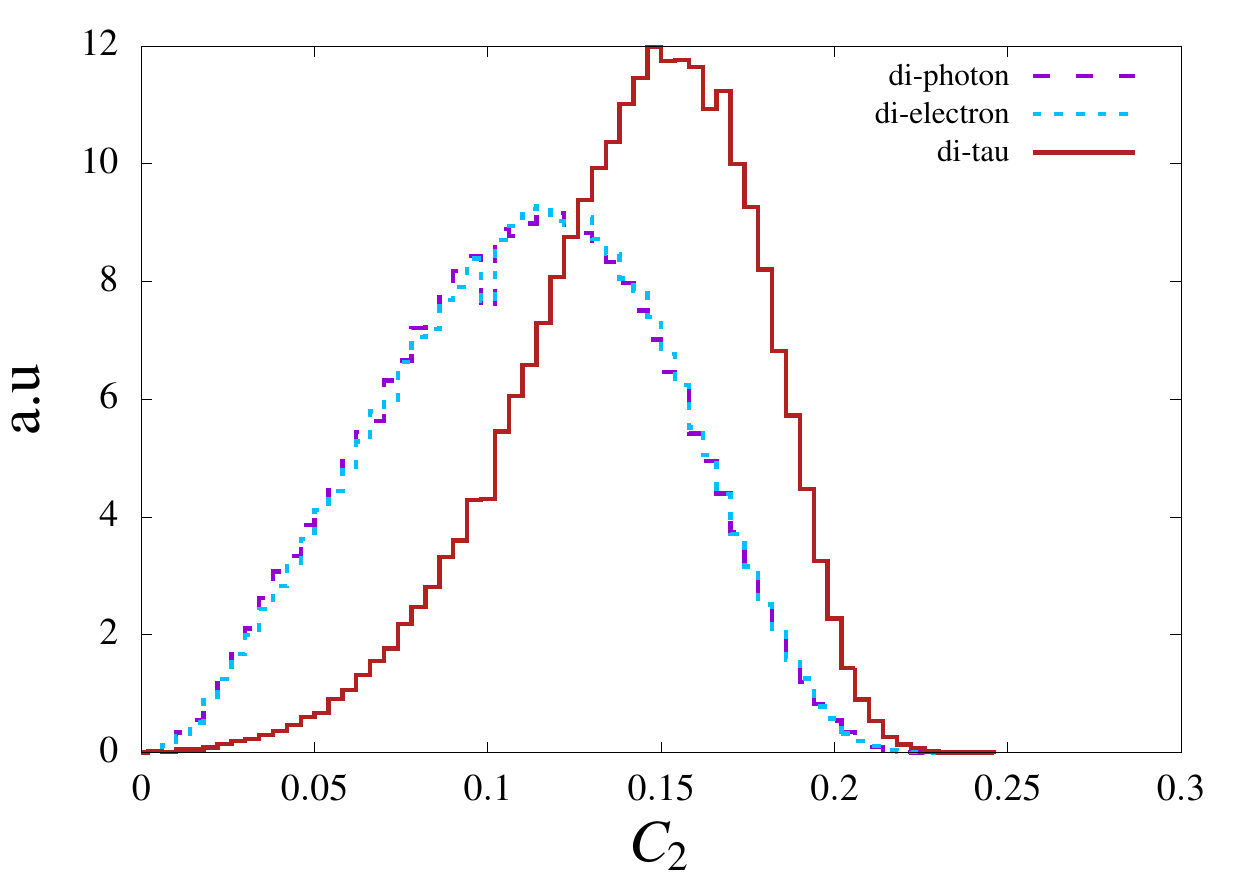}\\
                \end{tabular}
        \end{center}
        \caption{Distribution of one of the ratios and double ratios of Energy Correlation Function (ECF) $r_2$ (left) and $C_2$ (right) respectively for all the non-standard objects. }
        \label{sig:ecf2}
\end{figure}
\item \underline{Energy-Correlation functions (ECFs) and ratios}: The key feature of this variable is that it quantifies the distribution of the energy inside a jet utilizing the information of the jet constituents. It is thus a direct probe of the pronginess of the jet. In Fig.\ref{sig:ecf1} we show the distribution for the two ECF variables, namely $e_2$ (left) and $e_3$ (right) (see Eq.~\eqref{eq:ecf} for definition) for the non-standard objects. As already discussed in Sec.\ref{sec:3}, the $e_{n+1}$ computed for a jet with $n$-energetic prongs is always suppressed w.r.t. $e_n$. Now, as all the non-standard objects are primarily two-pronged, $e_3$ is expected to peak at much lower values compared to $e_2$. We validate our expectation in Fig.\ref{sig:ecf1} where one can indeed see the distribution of $e_3$ is left-shifted (towards lower values) compared to $e_2$. A similar feature can be observed in Fig.\ref{sig:ecf2} where we plot the ECF ratios $r_2 = e_3/e_2$ (left) and $C_2=e_3 e_1/e_2^2$ (right). Since $e_2$ is always greater than $e_3$, $r_2$ peaks at values $\le$ 1 for all the non-standard objects. The larger values of $r_2$ can be understood from the long tail in the $e_3$ and $e_2$ distributions. It is interesting to note that $C_2$ has some discrimination power for the di-tau jets from the other two di-samples. This can be attributed to the slight difference observed in the peak (and tail) of $e_2$ and $e_3$ distributions for di-tau jets. These minor differences get accentuated and, thereby the peak for the di-tau samples get shifted towards slightly higher values.

\end{itemize}

\begin{figure}[h]
        \begin{center}
         	\includegraphics[width=\textwidth]{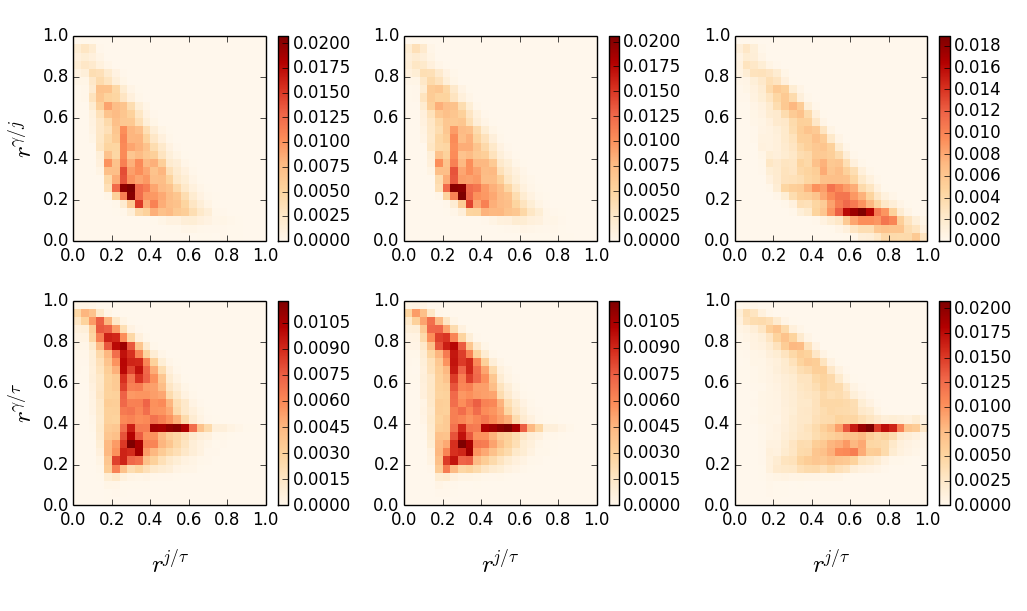} \\
        \end{center}
        \caption{ \label{fig:Responses_2d_anomalous} The 2-dimensional distributions of the BDT response variables $\rgj$,  $\rjt$, and $\rgt$ for the anomalous jets. The columns from the left to right represent the distributions for the di-photon, di-electron, and di-$\tau$ jets respectively. In these plots we have used $2D$-bins of size $\left(0.04\times 0.04\right)$ in units of responses.}
\end{figure}
It is clear that the segmentation of phase space already separates these three kinds from each other and identify their potential backgrounds. The di-photon jets  mostly occupy EC0, di-electrons occupy mostly EC2, whereas  di-taus can be found from HC1 to HC3. Following  the guideline discussed in Sec.\ref{sec:2.3}, we find representations of the anomalous objects in three dimensions (given by the three BDTs as discussed in Sec.~\ref{sec:4.2}). We show the 2-dimensional distributions of the BDT response variables $\rgj$,  $\rjt$, and $\rgt$ for these anomalous objects in Fig.\ref{fig:Responses_2d_anomalous} where the three columns represent di-photons (left), di-electrons (middle), and di-tau (right) jets. 

\begin{table}[htb!]
        \begin{center}
                \begin{tabular}{|c||c||c|c|c|}
\hline
                        \multirow{2}{*}{Regions}  & \multirow{2}{*}{Vetoes used} &
                        \multirow{2}{*}{$\epsilon_{\gamma\gamma}$ (in $\%$)}&
                        \multirow{2}{*}{$\epsilon_{ee}$ (in $\%$)}&
                        \multirow{2}{*}{$\epsilon_{\tau\tau}$ (in $\%$)} \\ &&&&\\
\hline
                        \multirow{2}{*}{EC0}  &
                        \multirow{2}{*}{Photon Veto with $R_{\gamma} = 0.05$} &
                        \multirow{2}{*}{59.6}&
                        \multirow{2}{*}{0.96}&
                        \multirow{2}{*}{0.21} \\ &&&&\\

                        \multirow{2}{*}{EC1}  &
                        \multirow{2}{*}{Electron Veto with $R_{e} = 0.05$} &
                        \multirow{2}{*}{17.1}&
                        \multirow{2}{*}{11.8}&
                        \multirow{2}{*}{1.4}  \\ &&&&\\

                        \multirow{2}{*}{EC2}  &
                        \multirow{2}{*}{No Veto} &
                        \multirow{2}{*}{2.8}&
                        \multirow{2}{*}{76.3}&
                        \multirow{2}{*}{2.5}  \\ &&&&\\
                        
                        \multirow{2}{*}{EC3+}  &
                        \multirow{2}{*}{No Veto } &
                        \multirow{2}{*}{0.83}&
                        \multirow{2}{*}{4.27}&
                        \multirow{2}{*}{0.48}   \\ &&&&\\
                        
\hline \hline
                        \multirow{2}{*}{HC0}  &
                        \multirow{2}{*}{QCD Veto with $R_{j} = 0.005$} &
                        \multirow{2}{*}{1.3}&
                        \multirow{2}{*}{0.04}&
                        \multirow{2}{*}{0.99}  \\ &&&&\\

                        \multirow{3}{*}{HC1}  &
                        \multirow{2}{*}{QCD Veto with $R_{j} = 0.005$} &
                        \multirow{3}{*}{0.21}&
                        \multirow{3}{*}{0.1}&
                        \multirow{3}{*}{2.7}         \\    
                   
                         & \multirow{2}{*}{Tau Veto with $R_{\tau} = 0.05$} &&&\\	 
                         &&&&\\

                        \multirow{2}{*}{HC2}  &
                        \multirow{2}{*}{QCD Veto with $R_{j} = 0.005$} &
                        \multirow{2}{*}{0.19}&
                        \multirow{2}{*}{0.42}&
                        \multirow{2}{*}{9.3}  \\ &&&&\\

                        \multirow{2}{*}{HC3}  &
                        \multirow{2}{*}{QCD Veto with $R_{j} = 0.005$} &
                        \multirow{2}{*}{0.06}&
                        \multirow{2}{*}{0.02}&
                        \multirow{2}{*}{2.86}   \\ &&&&\\
                        
                        \multirow{2}{*}{HC4+}  &
                        \multirow{2}{*}{QCD Veto with $R_{j} = 0.005$} &
                        \multirow{2}{*}{0.01}&
                        \multirow{2}{*}{0}&
                        \multirow{2}{*}{1.79}  \\ &&&&\\

\hline
                \end{tabular}
                \caption{The numbers represent the fraction of non-standard jets remain after the vetoes have been imposed.}  
                \label{table:eff_vetoes_non-standard}
        \end{center}
\end{table}

In Table~\ref{table:eff_vetoes_non-standard}, we summarize the effect the SM vetoes (as tabulated in Table~\ref{table:eff_vetoes}) 
on the non-standard di-samples. The numbers denote the fraction of the anomalous objects remain after the vetoes have been imposed in both the ECAL and HCAL regions segmented with different track multiplicities. The di-photon and di-electron jets are conspicuous by their presence in the EC0 and EC2 regions respectively. The photon veto with $R_{\gamma} = 0.05$ selects events with the leading jet having two or more photons. Thus, we observe relatively low yield ($\sim$ 5\%) for the single photon samples in the EC0, however a large yield of 60\% for the di-photon samples. Similar arguments hold for the di-electron samples in the EC2 region, where single electron and $\tau$ jets yields are 3.8\% and 0.44\% respectively in comparison to 76\% yield for the di-electron. In EC1 we observe a higher efficiency for the di-photon which can be attributed to the fact that one of the photons can get converted to an electron-positron pair with one of them showing up in the tracker. The di-tau sample with both the taus decaying hadronically is expected to have relatively lower yields in the ECAL regions with varying track multiplicities, and thus mild sensitivity is observed for the photon and/or electron vetoes. It is worth mentioning that EC3+ being supposedly free from the standard objects has a significantly larger efficiency for the di-electron and somewhat milder ($\sim 1\%$) efficiency for the di-photon and di-tau samples.

The single tau and QCD-jets constitute the major background in the hadronic calorimeter region. A QCD veto with $R_{j} = 0.005$ is imposed for all the segments irrespective of the track multiplicities. The di-tau jets, which is predominantly composed of two tracks has the maximum acceptance in the HC2 region with an efficiency of 9.3$\%$ with an acceptance rate of 2.3$\%$ and 0.55$\%$ for the single tau and QCD-jets respectively. Segments with one and three tracks (HC1 and HC3) also provide an appreciable amount of sensitivity for the di-tau samples when for HC1 a tau veto is additionally imposed. It is interesting to note that the cumulative percentage from the HC1 to HC3 for the di-tau sample is characterized by an acceptance of $16.7\%$, while for the single tau and QCD are 12.6$\%$ and 2$\%$ respectively. One may be worried about the yield of di-tau to be comparable to that of a single tau jet, however note that the vetoes were developed by adapting an approach of being agnostic of any non-standard physics. Furthermore, production rate for these single tau events are also much smaller compared to the QCD-jets. Thus, once events with a hint of di-tau signals are triggered, one may repeat the analysis by optimizing the separation of the di-tau jets from single tau and QCD jets as demonstrated for example in \cite{Englert:2011iz}. 

To summarize, in this subsection we demonstrate examples of anomalous objects (collinear particles) passing vetoes that restrict all standard objects (below a pre-determined acceptance rate). Even though the vetoes are constructed without using any information about the anomalous objects, we manage of find anomalous objects at a reasonable rate. 

\section{\label{sec:6}Conclusions}
The hunt for new physics constitutes an essential ingredient for the current and future run of the LHC. A fundamental assumption employed in these searches is that any new physics is characterized in terms of the standard reconstructed objects, such as isolated photons, electrons, taus,  QCD-jets etc. This strategy fails when new physics, instead, give rise to anomalous objects, such as collimated and equally energetic particles, or particles with long lifetime, to name a few.  These objects either are missed or are mis-identified as standard-objects. In case these are missed, we lose events unless associated particles trigger. In case, these are mis-identified, we mischaracterize the full event information.  Specifically, if we mis-identify these objects as QCD-jets the event gets lost in the sea of SM events due to QCD. Various studies have been proposed towards the discovery of these anomalous objects. However, proposals, typically, rely heavily on specifics of the anomalous objects themselves, which implies that these methods may lose sensitivity fast even for slightly altered NP scenarios.  

In this work we propose a framework where we identify these anomalous objects  entirely by constructing vetoes around the standard objects. The occurrence of an object passing all vetoes signify the detection of anomalous objects, which, in turn, gives hint of NP.  The framework for constructing vetoes as proposed here rely on, $(i)$ the use of jet-clustering algorithms as a universal construct for all objects (standard or non-standard),  $(ii)$ an ensemble of conventional and jet-substructure variables to find representations of jets in a multi-dimensional space, $(iii)$ the combination of phase-space segmentation and MVAs to reduce the dimensionality of the space without sacrificing information pertaining to pairwise differences among standard-objects, and finally $(iv)$ an algorithm (loosely based on the greedy algorithm) to identify regions rich in standard-jets. The procedure proposed here is completely agnostic of the form of new physics and therefore can be widely applied across different new physics scenarios which may give rise to such anomalous objects.

Notice that the current set up of the proposed ``Anomaly Finder" does not include 
the Muons and b-jets. The identification and reconstruction of Muons and
b-jets at the LHC involve specialized techniques. In the existing set up, the b-jets would fall into the category of
identified QCD-jets. However, note that b-jet reconstruction strategy at the LHC includes the combined information of the calorimeter energy
deposits as well as information of displaced tracks and properties of secondary and tertiary decay vertices
reconstructed within the jet \cite{Sirunyan:2017ezt,Aad:2015ydr}. These additional information will 
thus introduce a collection of new kinematic variables, especially
in terms of vertex and life-time information of the B-hadrons. The inclusion of this information in the proposed framework
is indeed interesting and a straightforward extension of the proposed framework. The Muons, at the LHC, are
reconstructed from the tracks in the inner detector and muon spectrometer information, which are then
combined to improve the reconstruction efficiency and background rejection rates \cite{Sirunyan:2018fpa,Aad:2016jkr}. 
Moreover, the Muon candidates are also required to satisfy stringent lepton isolation cuts. In this work, we reconstruct
jets using calorimeter information only, and so we don't have the full information for the Muons. However, we can still define a region of
parameter space which should be Muon-rich. For example, we can look for events with exactly one track
associated to the jet with negligible energy depositions both in ECAL and HCAL. This segment of phase-space is
very unique, and has almost no overlap with $\tau$- or QCD-rich jets. Additionally, we can also extend the
existing set up incorporating variables based on Muon spectrometer information.

Before we end, a practical guideline on the implementation of this proposal is worth mentioning. 
Here we propose two strategies to categorize the data samples to be analyzed at the LHC. First one is an 
{\it offline} analysis, while the other an {\it online} implementation. The offline mode assumes that the 
event has already been triggered at the HLT level through the existing trigger menu by reconstructing, 
for each event, objects like electrons, muons, and jets and then selected based on several identification criteria and 
physics related goals. Once the events are triggered and selected, the proposed analysis, namely 
the `anomaly finder', can be performed independently to look for new physics signatures. Of course, here 
we assume that the anomaly finder has already been optimized using the control sample, and thus one needs 
to simply pass the registered events through the anomaly finder. Further, one can also 
use additional information from the processes like the associated production of Higgs boson with a Z-boson (with Z decaying to 
muons or invisibly), or say pair produced Z-bosons with both Z decaying leptonically etc., to model 
the standard objects in the Higgs or, Z channels. Here we stress that all of these analyses can be performed offline, 
and thus, this proposal provides a unique framework to probe a wide range of new physics scenarios by directly identifying 
events containing anomalous objects. Note that, one can always perform supervised analysis later to probe the 
origin and nature of those anomalous objects. 

The second approach, a bit more aggressive, is to combine the proposed `anomaly finder' with the existing HLTs, which 
will provide a unified framework to look for direct imprints of new physics in the LHC data. It is interesting to note 
that both the ATLAS and CMS collaborations at the LHC have modified and redesigned the trigger menu significantly to 
cope with the higher event rates at run-2 as well as high luminosity runs of LHC \cite{Aaboud:2016leb,Khachatryan:2016bia}. 
The HLT softwares are now upgraded to enhance the acceptance rates by making the algorithms and selections criteria similar to the 
offline reconstruction techniques for objects like electrons, muons and jets. Interestingly, anti-$k_T$ jets with varying values 
of jet radius are reconstructed at the HLT with the calorimeter topo-clusters constructed from the calorimeter cells. These jets 
are then calibrated for the nonlinearity of the calorimeter response and pileup effects using a combination of 
studies based on simulation and collision data. Identification and tagging the flavour of these reconstructed jets, 
e.g. b-jet tagging, tau-tagging etc., are now an integral part of the HLT system. Moreover, these updated online flavour tagging 
templates now include advanced multivariate analysis (MVA) incorporating various discriminating variables mimicking their 
offline templates \cite{ATL-PHYS-PUB-2017-013,Chatrchyan:2012jua,ATLAS:2017mpa,Sirunyan:2017khh}. Search 
for exotic new physics signatures at the LHC, for example, long-lived particles, displaced jets, displaced 
leptons etc., also utilize sophisticated MVA-based techniques and algorithms especially deigned to trigger these rare 
events, for example \cite{CMS:2016dla}. Thus, we understand that the existing HLT set up is already efficient enough 
to handle sophisticated algorithms similar to their offline counterparts, and provide impressive results. 
The proposed `anomaly finder' require to construct several variables utilizing the tracker and calorimeter 
information, and perform a MVA to obtain a collection of vetoes that eliminate all standard-objects upto a 
pre-determined acceptance rate. In this work we assume the acceptance rate for the QCD-jets to 
be $0.5\%$, while the existing HLT photon trigger menu accepts isolated photons ($p_{T} >$ 20 GeV, loose selection) 
with an efficiency of 97\% with a rejection factor for the QCD-jets around 1000 \cite{ATLAS:2011kuc,Aaboud:2016yuq}. A 
crucial aspect of the proposed anomaly finder is that it includes a free/input parameter that directly 
controls the rate at which QCD jets get accepted. Our choice was essentially aimed to provide a 
concrete example, however one can always tune the parameter associated to the QCD rejection rate to a desired value 
while probing a wide class of new physics models.    

Therefore, this proposal can be used either as a stand-alone framework (offline mode) 
once we select events after the HLT with acceptable event rates, or we combine it with the existing HLT menu (online mode) 
with moderate thresholds for the SM event rates. Both the strategies are expected to work reasonably well with 
the real data.

\newpage
\section*{Acknowledgements}

We thank Adam Martin and Michael Graesser for careful reading of an earlier version of the draft and sending critical remarks. A significant part of the computations was completed in the Gaggle cluster at TIFR. Some preliminary simulations were also carried out at the Mapache cluster in the HPC facility at LANL. TSR was supported by the Early Career Research Award by Science and Engineering Research Board, Dept. of Science and Technology, Govt. of India (grant no: ECR/2015/000196). We also thank Sreerup Raychaudhuri for helping us with computational resources.

\appendix

\section{\label{sec:appA} Simulation Details}

\begin{enumerate}

\item {\underline {Event simulation}}: As outlined in Sec.\ref{sec:2.2}, standard-jets of various kinds (i.e., single photon, single electron, single tau, QCD-parton initiated jets) are constructed from the leading jet of an event. For example, photon (or jet of type photon) is the leading jet in events with $pp \rightarrow h \rightarrow \gamma \gamma$ where $h$ represents the 125 GeV SM Higgs boson. The event generation as well as parton showering and hadronization have been performed using \texttt{Pythia 8.2}~\cite{Sjostrand:2007gs} with parton distribution function \texttt{NNPDF2.3} \cite{Ball:2013hta}. For the non-standard jets, we implement the toy Lagrangian described in Eq.~\eqref{eq:toy_lagrangian1} in \texttt{FeynRules 2.0}~\cite{Christensen:2008py}. The generated model files are then used to generate the events using \texttt{MadGraph 2.3.3}~\cite{Alwall:2014hca}. The events are then passed to \texttt{Pythia} for showering and hadronization. 

\item {\underline {Detector simulation}}: In order to perform a fast detector simulation, we use \texttt{Delphes 3.3.2} \cite{Ovyn:2009tx,deFavereau:2013fsa} with the CMS card. The default charged and neutral particle identification efficiencies as implemented in the card have been used. We simulate low-$Q^2$ soft QCD pile-up events using {\tt Pythia} and then pass it through {\tt Delphes}. The default parametrization as implemented in the CMS card has been used to distribute the minimum-bias pile-up events and hard scattering events in time and $z$ positions. The mean number of soft events merged with each hard scattering, denoted by $\langle N_\text{PU}\rangle$, is considered to be $40$. Note that, after adding these low-$Q^2$ soft QCD events, one has to identify the primary vertex and then remove those collisions which are not associated to the primary vertex; one can achieve this by performing a pile-up subtraction technique. 

A combination of vertex and tracker information helps to identify (and then remove) the contamination of the charged particles originating from the pile-ups. On the other hand, contribution of neutral particles to the pile-up events can be estimated, and then physical observables can be accordingly corrected, by using the jet area method \cite{Cacciari:2008gn,Cacciari:2007fd}. In this work, we follow the default set up of {\tt Delphes} CMS card to perform the pile-up subtraction. A spatial vertex resolution parameter $|z|$ is used to perform the charged pile-up subtraction; every charged particle originating from a reconstructed vertex with $|z| > $ 0.01 cm are considered as coming from pile-ups. We consider those tracks which are passed through the \texttt {TrackPileUpSubtractor} module in {\tt Delphes}. Jets are constructed with the calorimeter tower elements using \texttt{Fastjet 3.1.3} \cite{Cacciari:2011ma} with anti-$k_T$ jet algorithm \cite{Cacciari:2008gp}, jet radius $R = 0.4$ with $p_T > 50\gev$. Similar to the tracks, we require to correct the reconstructed jets from low-$Q^2$ pile-up events containing neutral particles. Note that, charged particles that have failed to be reconstructed as tracks or, are outside the tracker volume can also contribute here. In {\tt Delphes}, the residual pile-up subtraction is achieved by using an algorithm based on the jet area. This technique helps to correct the jet momenta by calculating pile-up density ($\rho$) and jet area. Here we use the jets constructed using the calorimetric information and allow the default estimation of $\rho$ with the {\tt EFlow} elements. Finally, we recluster the constituents of the pile-up subtracted leading jet ($p_T$ ordered), obtained from the \texttt {JetPileUpSubtractor} module, to find an exclusive C/A jet \cite{Dokshitzer:1997in}. This pile-up subtracted C/A jet is considered in rest of our analysis. The last step of jet clustering is performed just to have a C/A-based clustering history of the jet. The number of tracks associated to the leading jet is counted by calculating $\Delta R$ between the jet and each pile-up subtracted track, and then accept those tracks with $p_T \geq 2\gev$ and $\Delta R <0.4$, where $\left(\Delta R\right)^2 \equiv \left( \Delta\eta \right)^2 + \left( \Delta\phi \right)^2$ with $\Delta\eta$ and $\Delta\phi$ being the differences in pseudo-rapidity and azimuthal angle between them respectively.

\item {\underline {Photon conversion}}: In order to implement conversion of photons in the tracker portion of the detector we simply follow the prescription as described in Ref.~\cite{Ellis:2012sd,Ellis:2012zp}. We register a track for photons after drawing a random number from $0$ to $1$ in a flat grid. The probability of conversion is $\eta$-dependent, since the amount of material a photon passes through (\textit{i.e.},  the number of radiation lengths) varies with directions. For simplicity, in this analysis we assign a flat conversion probability of 20\%.

\item {\underline {BDT parameters}}: The parameters associated to BDT analyses are chosen as follows:  the number of trees in the forest {\tt NTrees}= $800$, the maximum depth of the decision tree {\tt MaxDepth} $=3$,  and finally, the minimum percentage of training events required in a leaf node {\tt MinNodeSize}  $=2.5\%$. All other necessary variables are kept at their default values. Furthermore, we consider the {\tt AdaBoost} 
method \cite{Freund:1997xna} for boosting the decision trees with the boost parameter {\tt AdaBoostBeta} $=0.5$.

\end{enumerate}

\bibliography{references}
\bibliographystyle{elsarticle-num}
 		
\end{document}